\documentclass[useAMS,usenatbib]{mn2e}

\pdfoutput=1

\usepackage{ifpdf}
\usepackage{amssymb,amsmath}
\usepackage{graphics}
\usepackage{subfigure}
\usepackage[dvips]{graphicx}
\usepackage{textcomp}

\def\ltsima{$\; \buildrel < \over \sim \;$}
\def\simlt{\lower.5ex\hbox{\ltsima}}   
\def\gtsima{$\; \buildrel > \over \sim \;$}
\def\simgt{\lower.5ex\hbox{\gtsima}}
\newcommand\bcite[1]{\citeauthor{#1} \citeyear{#1}}

\newcommand{\hi} {{\rm H}\,{\small\rm I}}
\newcommand{\kms} {\,{\rm km\,s}^{-1}}
\newcommand{\cm} {\,{\rm cm}^{-3}}

\newcommand{\kpc} {\,{\rm kpc}}
\newcommand{\Mpc} {\,{\rm Mpc}}

\newcommand{\mo}{\,{M}_\odot}

\newcommand{\Myr}{\,{\rm Myr}}
\newcommand{\Gyr}{\,{\rm Gyr}}
\newcommand{\K}{\,{\rm K}}
\newcommand{\erg}{\,{\rm erg}}

\newcommand{\mopc}{\,M_{\odot} {\rm pc}^{-2}}
\newcommand{\moyr}{\,M_{\odot} {\rm yr}^{-1}}
\newcommand{\moyrkpc}{M_\odot {\rm yr}^{-1} {\rm kpc}^{-2}}

\newcommand{\gsim}{\lower.7ex\hbox{$\;\stackrel{\textstyle>}{\sim}\;$}}
\newcommand{\lsim}{\lower.7ex\hbox{$\;\stackrel{\textstyle<}{\sim}\;$}}

\def\ncor{n_{\rm cor}}
\def\ncorp{n_{\rm cor, p}}
\def\ncormin{n_{\rm cor}|_\mathrm{min}}
\def\ncormax{n_{\rm cor}|_\mathrm{max}}
\def\ncorminthree{n_{\rm cor,3}|_\mathrm{min}}
\def\ncorminone{n_{\rm cor,1}|_\mathrm{min}}

\title[Unveiling the Milky Way's corona]
{Unveiling the corona of the Milky Way via ram-pressure stripping of dwarf satellites}

\author[A. Gatto, F. Fraternali, J. I. Read, F. Marinacci, H. Lux and S. Walch]
{A. Gatto,$^{1,2}$\thanks{e-mail: andreag@mpa-garching.mpg.de}
F. Fraternali,$^{2,3}$ J. I. Read,$^{4}$ F. Marinacci,$^{5,6}$ H. Lux$^{7,8}$ and S. Walch$^{1}$\\
$^{1}$Max-Planck Institute for Astrophysics, Karl-Schwarzschild Strasse 1, D-85748 Garching, Germany\\ 
$^{2}$Department of Physics and Astronomy, University of Bologna, via Berti Pichat 6/2, I-40127 Bologna, Italy\\
$^{3}$Kapteyn Astronomical Institute, Postbus 800, NL-9700 AV, Groningen, The Netherlands\\
$^{4}$University of Surrey, Guildford, Surrey GU2 7XH, UK\\
$^{5}$Heidelberg Institute for Theoretical Studies, Schloss-Wolfsbrunnenweg 35, 
D-69118 Heidelberg, Germany \\
$^{6}$Center for Astronomy of Heidelberg University, Astronomisches Recheninstitut, 
M\"onchhfstr. 12-14, D-69120 Heidelberg, Germany \\
$^{7}$School of Physics and Astronomy, University of Nottingham, University Park, Nottingham NG7 2RD, UK  \\
$^{8}$Department of Physics, University of Oxford, Denys Wilkinson Building, Keble Road, Oxford OX1 3RH, UK}

\begin{document}

\date{Accepted 2013 May 17. Received 2013 May 16; in original form 2013 February 25}

\pagerange{\pageref{firstpage}--\pageref{lastpage}} \pubyear{2013}

\maketitle

\label{firstpage}

\begin{abstract}
The spatial segregation between dwarf spheroidal (dSph) and dwarf
irregular galaxies in the Local Group has long been regarded as
evidence of an interaction with their host galaxies. In this paper, we
assume that ram-pressure stripping is the dominant mechanism that
removed gas from the dSphs and we use this to derive a lower
bound on the density of the corona of the Milky Way at large
distances ($R\sim 50-90 \kpc$) from the Galactic Centre. At the
same time, we derive an upper bound by demanding that the interstellar
medium of the dSphs is in pressure equilibrium with the hot corona.
We consider two dwarfs (Sextans and Carina) with well-determined
orbits and star formation histories. Our approach introduces several
novel features: (i) we use the measured star formation histories of
the dwarfs to derive the time at which they last lost their gas, and
(via a modified version of the Kennicutt-Schmidt relation) their
internal gas density at that time; (ii) we use a large suite of 2D
hydrodynamical simulations to model the gas stripping; and (iii) we
include supernova feedback tied to the gas content. Despite having
very different orbits and star formation histories, we find results
for the two dSphs that are in excellent agreement with one another. We
derive an average particle density of the corona of the Milky Way at
$R=50-90 \kpc$ in the range $n_\mathrm{cor}=1.3-3.6 \times 10^{-4}
\cm$. Including additional constraints from X-ray emission limits and
pulsar dispersion measurements (that strengthen our upper
bound), we derive Galactic coronal density profiles. Extrapolating
these to large radii, we estimate the fraction of baryons (missing baryons) that can
exist within the virial radius of the Milky Way. For an isothermal
corona ($T_{\rm cor}=1.8\times 10^6 \K$), this is small -- just
$10-20\%$ of the expected missing baryon fraction, assuming a virial mass of
$1-2 \times 10^{12} \mo$. Only a hot ($T_{\rm cor}=3\times 10^6 \K$)
and adiabatic corona can contain all of the Galaxy's missing baryons.
Models for the Milky Way must explain why its corona is in a hot
adiabatic thermal state; or why a large fraction of its baryons lie
beyond the virial radius.
\end{abstract}


\section{Introduction}\label{sec:intro}

In the current cosmological framework, the fraction of baryonic matter to
dark matter (DM) is known to a high level of precision, thanks to both big-bang
nucleosynthesis \citep{Pagel97} and the study of the cosmic microwave background
\citep[e.g.][Planck Collaboration \citeyear{PlanckColl13}]{Komatsu09}. By contrast, the fraction of
baryons observed in the form of stars and gas in collapsed structures in the
Universe is rather scant, which is commonly referred to as the {\it missing
baryon problem}. Only massive galaxy clusters appear to have the amount of
baryons expected, mostly in the form of hot gas that permeates their deep
potential wells \citep[e.g.][]{Sarazin09}. Galaxy groups and isolated galaxies
contain a fraction of detectable baryons which is a factor of $\sim 10$ smaller than
the expected fraction and this discrepancy steadily increases with decreasing
virial mass \citep[e.g.][]{2005RSPTA.363.2693R,McGaugh+10}.

Disc galaxies represent particularly challenging environments. Applying the
cosmological baryon fraction to the virial mass of the Milky Way (MW; $1-2 \times10^{12}\rm\mo$, \citealp[e.g.][]{Wilkinson&Evans99}), 
one would predict a total baryonic mass for the Galaxy of $\sim 2-3 \times
10^{11}\rm\mo$. However, the currently detected mass in stars is $\sim 5 \times
10^{10}\rm\mo$ \citep{Dehnen&Binney98} while interstellar matter
accounts only for $< 1 \times 10^{10}\rm\mo$
\citep{Binney&Merrifield98,Nakanishi&Sofue06}. Therefore, $\sim 70-80$\% of
the MW's baryons are missing. Similar discrepancies are obtained
for other disc galaxies of comparable mass \citep[e.g.][]{2005RSPTA.363.2693R}.

A commonly accepted solution to this incongruity is that galaxies should be
embedded in massive atmospheres -- {\it cosmological coronae} -- of hot gas at
temperatures of a few $10^{6}$\,K which contain most of the baryons associated
with their potential wells \citep{Fukugita&Peebles06}. To date, the detection of
these coronae has proven rather elusive since at this temperature and density
(and assuming a low metallicity) the gas is unable to efficiently absorb or emit
photons through metal lines or bremsstrahlung radiation
\citep[e.g.][]{Sutherland93}. Some disc galaxies do show X-ray emission outside
of their discs, but in most cases this is clearly associated with star formation
and the presence of galactic winds \citep{Strickland+04}. In general, owing to
contamination from the disc, an unambiguous detection of a cosmological corona
is difficult in disc galaxies, unless hot gas is seen at large distances
($\simgt 10 \kpc$) above or below the disc plane. A notable case is the
massive galaxy NGC\,5746 \citep{Pedersen+06}, where an early claim of an
extended X-ray emitting corona was later attributed to an error in the
background subtraction in the {\it Chandra} data \citep{Rasmussen+09}. This case
alone demonstrates that these studies are at the limit of the capabilities of
current X-ray facilities \citep[]
[but see also \citealt{Hodges-Kluck&Bregman13,LiWang12}]{Bregman07}.

In the MW, there are several indirect indications of the presence of a
hot corona. The first evidence was pointed out by Spitzer soon after the
discovery of clouds at high latitude as a medium capable of providing their
pressure confinement \citep{Spitzer56}. Head-tail shapes of high-velocity clouds (HVCs) are also
considered as evidence of an interaction between them and the corona
\citep{Putman+11}. Unfortunately, all measured distances of HVCs are within $10
\kpc$ from the plane of the disc \citep[e.g.][]{Wakker+08}. Thus it is not clear
whether they are probing the cosmological corona or simply extra-planar hot gas. A
perhaps more relevant observation is the asymmetry between the leading and
trailing arms of the Magellanic Stream \citep[e.g.][]{Putman+03b}, which is
seen further out ($\sim 50 \kpc$) and could result from ram-pressure stripping
\citep[see][]{Guhathakurta&Reitzel98,Mastropietro+05,DiazBekki12}. 
Finally, X-ray spectra towards bright active galactic nuclei (AGN) show
absorption features -- in particular O\,{\small VII}, Ne\,{\small IX}, and
O\,{\small VIII} -- characteristic of a corona at $T \gsim 10^6 \K$. However,
the poor velocity resolution of these spectra does not allow us to determine the
extent of this medium and the current estimates range from a few kpc to $\sim 1
\Mpc$ \citep{Nicastro+02, Bregman&Lloyd-Davies07, Yao+08}.

\citet[][hereafter AB10]{Anderson&Bregman10} list a number of known indirect
pieces of evidence for the Galactic corona. They attempt to use them to give
limits on the amount of gas it can contain. For an isothermal corona,
they argued that the gas mass should be relatively small -- of the order of
10\% of the total mass of missing baryons -- assuming a \citet{NFW} (NFW) profile.
The fraction can become significantly larger, however, for {\it adiabatic}
coronae \citep[see also][]{Binney+09, Fang+12}. The same authors presented also a
possible detection of a corona of missing baryons around the massive spiral NGC\,1961
\citep{Anderson&Bregman11}. Their estimate of the total mass for an isothermal
corona is again $\sim 10$ \% of the baryons that should be
associated with the potential well of this galaxy. This estimate comes from an
extrapolation as the visible corona extends only to about $\sim50 \kpc$ from the
centre. Potential problems with this detection come from the fact that this
galaxy may be the result of a recent collision \citep{Combes+09} and shows a
rather disturbed \hi\ disc that extends to a distance of 50\,kpc from the centre
\citep{Haan+08}. A new and more compelling detection is that of the
super-massive disc galaxy UGC\,12591, where the amount of gas in the corona is
estimated to be between 10\% (isothermal) and 35 \% \citep[adiabatic;][]{Dai+12}.

Following \citet{Shull+12}, the low-redshift baryon content can be divided as
follows: 1.7\% in cold gas (HI and HeI), 4\% in the ICM (Intra-Cluster Medium), 5\% in the CGM
(Circum-Galactic Medium)\footnote{Due to the poor knowledge of such a phase,
this fraction has been assumed rather than measured.}, 7\% in galaxies (stars and ISM), 30\% in
the intergalactic WHIM (Warm-Hot Ionised Medium) and 30\% in the Ly$\alpha$ forest. 
This leaves $29\pm13$\% of the baryons still missing.
From their high-resolution cosmological simulations, these authors found that about half of these
missing baryons may be in a hot ($T>10^6$ K) intergalactic WHIM phase.

In this paper, we derive the density of the corona of the Mw at large
distances ($\sim 50-90 \kpc$) from the centre using the population of surrounding
dwarf spheroidal (dSph) satellites as a probe of the hot halo gas. 
dSphs are gas-free dwarf galaxies -- at least down to current detection
limits \citep[e.g.][]{Mateo98}. 
They are typically located close to their host galaxy in contrast to the gas-rich dwarf
irregulars (dIrrs) that lie at larger distances \citep{Mateo98, 2006ApJ...653..240G}. 
The proximity to our Galaxy is believed to be the reason for the removal of material from
the dSphs, as several other physical properties are very similar between the two types \citep[e.g.][]{Kormendy85, Tolstoy+09}. 
A similar {\it distance-morphology relation} is also observed in dwarf galaxies in other groups \citep[e.g.][]{2006ApJ...653..240G}, suggesting that in addition to supernova (SN) feedback,
environmental effects like ram-pressure stripping from a hot corona \citep{Gunn&Gott72, Nulsen82} or
tidal stripping \citep[e.g.][]{2006MNRAS.366..429R,2006MNRAS.367..387R} must play a crucial role. 
There is a vast literature investigating these phenomena via hydrodynamical
simulations in different environments, from galaxy clusters to MW-sized haloes
\citep[e.g.][]{Mori&Burkert00, Marcolini+03, Roediger&Hensler05,
Nichols&Bland-Hawthorn11}, as well as observations of possible on-going ram-pressure stripping from dwarf galaxies \citep{2007ApJ...671L..33M} and normal galaxies \citep{2012A&A...544A.128F}.
For a study that combines ram-pressure and tidal stripping in dwarfs see \citet{Mayer+06}.

Here, we concentrate on ram-pressure stripping and assume it to be the dominant
mechanism that removed gas from the dSphs (tidal stripping plays a more 
minor role for the galaxies we study here; see \S\ref{sec:tides} and \citealt{BlitzRobishaw00}). We introduce a
simple model of SN feedback and investigate its influence on the stripping rate. We then
estimate the minimum density that the corona of the MW should have for
this stripping to occur. This technique has been pioneered by \citet{LinFaber83,MooreDavis94}
and subsequently refined by \citet[and see also \citealt{BlitzRobishaw00}]{Grcevich&Putman09}, who considered a simple analytical formula for the stripping, applied it to four
dSphs, and found that the number density of the hot halo within $\sim 120$\,kpc
from the centre of the MW is of the order of a few times $10^{-4} \cm$. 
In this paper, we improve on these earlier works by adding several novel features: 
\begin{enumerate} 
\item we perform hundreds of 2D hydrodynamical simulations of gas stripping;
\item we use the measured star formation histories (SFHs) for the dwarfs to derive the time at which they last lost their gas and, using a modified version of the Kennicutt-Schmidt
(K-S) relation \citep{Schmidt59, Kennicutt98a}, we determine their internal gas density at that time;
\item we use a detailed reconstruction of the orbits of the dwarfs that fully marginalizes over uncertainties in their distances, line-of-sight velocities and proper motions;
\item we include a model for SN feedback with discrete energy injections to assess the importance of internal versus external gas loss mechanisms; and
\item we use pressure confinement arguments (similar to \cite{Spitzer56} but applied to the dSphs) to derive an {\it upper bound} on the coronal density.
\end{enumerate} 
We use the Sextans and Carina dwarfs, which are suitable for this study because they are small systems with reliable SFHs and mass estimates.
Moreover they have similar pericentric radii but totally different SFHs, providing an excellent consistency test of our method.

This paper is organized as follows. In \S\ref{sec:theory}, we estimate the effect of ram-pressure stripping on dwarf galaxies. We estimate the relative importance of tidal stripping for the two dSphs we study here, and we introduce the key concepts used in this paper to derive our lower and upper bounds on the coronal gas density.  In \S\ref{sec:method}, we describe our numerical method and initial conditions. In \S\ref{sec:results}, we show our results. In \S\ref{sec:discussion}, we wrap in other constraints on the coronal density from the literature and discuss the implications of our results for the missing baryon problem. Finally, in \S \ref{sec:conclusions} we present our conclusions.

\section{Analytic results}\label{sec:theory}

\subsection{Ram-pressure stripping: a lower bound on the hot corona density}\label{sec:ramp} 
To leading order, a dwarf galaxy will be ram-pressure stripped of its
ISM if \citep{Gunn&Gott72}: 
\begin{equation} 
\rho_{\rm cor} v^2 \gsim \rho_{\rm gas} \sigma^2\ , 
\label{eqn:crudestripform} 
\end{equation} 
where $\rho_{\rm cor}$ is the density of the background medium (the Galactic corona) that we would like to
measure, $v$ is the velocity of the dwarf galaxy, $\rho_{\rm gas}$ is the density of gas in the dwarf's ISM and $\sigma$
is the velocity dispersion of the dwarf (a proxy for its mass).

The velocity of the dwarf $v$ is maximized at the pericentre of the orbit, 
as is the background density $\rho_{\rm cor}$.
Thus, we can reasonably expect almost all of the ram-pressure stripping 
to occur at or near to a pericentric passage. 
At the orbital pericentre $r_{\rm p}$, equation (\ref{eqn:crudestripform})
can be recast as:
\begin{equation}
\left.\rho_{\rm cor}(r_{\rm p})\right|_\mathrm{min} = \frac{\rho_{\rm gas} \sigma^2}{v(r_{\rm p})^2}\ ,
\label{eqn:stripformperi} 
\end{equation} 
which gives the minimum coronal density at $r_{\rm p}$ required to strip 
the dwarf of all of its ISM, assuming that $\rho_{\rm gas}$ is the density of the latter just 
before the stripping event.

The velocity of the dwarf at the pericentre $v(r_{\rm p})$
and the pericentre value $r_{\rm p}$ are easily determined once the orbit is known.
Dwarf orbits can be reconstructed by assuming simple spherical potential 
models for the MW up to $\sim 2$ orbital periods backwards in time 
\citep{Lux+10}, and in some cases even more depending on how close to spherical 
the background potential is, and whether or
not the dwarf fell in isolation or inside a `loose group'. The velocity
dispersion of the dwarf $\sigma$ can be obtained from stellar kinematic
measurements \citep[e.g.][]{Walker+09}, which just leaves the ISM density
$\rho_{\rm gas}$ as a free parameter. A novel key aspect of this work is that we
introduce a new method for estimating $\rho_{\rm gas}$. Using deep resolved colour
magnitude diagrams, and fitting stellar population synthesis models, the SFH of the nearby MW dwarf
galaxies can be inferred \citep[e.g.][]{2005astro.ph..6430D}. This gives us the star formation
rate as a function of time from which we can derive the last moment at which the
dwarf had gas available to form fresh stars. Furthermore, through a modified version of the
K-S relation,
we can estimate the gas surface density at this time, $\Sigma_{\rm gas}$
(see \S \ref{sec:initialConditions}). Assuming
spherical symmetry and de-projecting we get the ISM density $\rho_{\rm gas}$.
All this information can then be used to solve equation (\ref{eqn:stripformperi})
for $\left.\rho_{\rm cor}(r_{\rm p})\right|_\mathrm{min}$.

In practice, we actually simulate the passage of a dwarf through pericentre in
order to retrieve more accurate results with respect to the analytic ones. The
simulations also allow us to include the effect of stellar feedback.
Equation (\ref{eqn:stripformperi}), however, remains useful as it captures the
essence of our methodology. We consider the accuracy of using equation (\ref{eqn:stripformperi})
as opposed to the full hydrodynamic simulations in \S \ref{sec:compAnalytic}.

\subsection{Pressure confinement of the dwarf ISM: an upper bound on the hot corona density}\label{sec:pconfine}
\begin{table*}
\centering
\begin{tabular}{lcccccccccc}
\hline
    dSph & Distance  & $L_V$     & $r_{\rm p}$ & $r_a$ & $t_\mathrm{orb}$ & $r_\mathrm{last}$ & $M_{\rm DM}(r_{\rm last})$ & $t_{\rm lb}$ \\
         & (kpc)         & ($10^6 L_\odot$)    & (kpc) & (kpc) & (Gyrs)             & (kpc)               & ($10^7 \rm\mo$)    & (Gyrs) \\
\hline
Sextans  & $86\pm4$ & $0.5$           & $60 \pm 20$ & $200 \pm 100$ & $4\pm 3$ & $1$ & $2$ & $\sim 7$ \\
Carina   & $101\pm5$& $0.43$         & $50 \pm 30$ & $110 \pm 30$ & $1.8\pm 0.8$ & $0.87$ & $3.7$ & $\sim 0.5$ \\
\hline
\end{tabular}
\caption{Physical properties of the Carina and Sextans dSph
galaxies. From left to right, the columns show the distance to the dwarf, the
$V$-band luminosity, the pericentre and apocentre, the orbital period, the
radius to the last measured kinematic data point $r_\mathrm{last}$, the mass
within $r_\mathrm{last}$, and the time to the last star formation burst $t_{\rm
lb}$. Data are taken from \citet{Mateo98} (distance, $L_V$), \citet{Walker+09}
($r_\mathrm{last}$, $M_{\rm DM}(r_{\rm last})$), \citet{Lux+10} ($r_{\rm p}$,
$r_a$, $t_\mathrm{orb}$). $t_{\rm lb}$ has been derived from the SFHs in
\citet{Lee+09} (Sextans) and \citet{Rizzi+03} (Carina).} 
\label{tab:properties}
\end{table*}

A novel idea in this work is to use the pressure confinement of the dwarf ISM to obtain an {\it upper bound} on the hot corona density \citep[c.f.][]{Spitzer56}. Matching the internal pressure of the dwarf ISM with the external pressure from the hot halo, we have:  
\begin{equation} 
\rho_{\rm gas} T_{\rm gas} \sim \rho_{\rm cor} T_{\rm cor}\ ,
\end{equation} 
where $T_{\rm gas} \sim 10^4$\,K is the temperature of the dwarf galaxy ISM, and $T_{\rm cor} \sim 10^6$\,K is the temperature of the MW hot corona. Thus, for a given total gas mass in the dwarf $M_\mathrm{gas}$, the dwarf ISM gas will extend to some maximum radius: 
\begin{equation}
r_\mathrm{gas} \sim \left[\frac{3M_\mathrm{gas}}{4\pi\rho_{\rm cor}}\frac{T_{\rm gas}}{T_{\rm cor}}\right]^{1/3}\ .
\label{eqn:rgas}
\end{equation}
We know $M_{\rm gas}$ from the SFH (see \S\ref{sec:ramp}) while $T_{\rm gas}$ and $T_{\rm cor}$ follow from our potential models for the dwarf and the MW. Thus, we can estimate $\rho_{\rm cor}$ simply from $r_\mathrm{gas}$. If we allow $r_\mathrm{gas}$ to extend to infinity, then we obtain essentially no bound on $\rho_{\rm cor}$. However, if we assume some minimum $r_\mathrm{gas}|_\mathrm{min}$, then we obtain an upper bound on the hot corona density $\rho_{\rm cor}|_\mathrm{max}$.

We assume here that $r_\mathrm{gas}|_\mathrm{min}$ is set by the radius within which the SFH history is derived ($r_\mathrm{SF}$, see later). This assumption is sensible since at the time of the last star formation event, the gas had to be at least as extended as the stars that formed from it. It is also self-consistent since $r_\mathrm{SF}$ is the radius out to which we estimate $M_\mathrm{gas}$. However, it relies on the stellar distribution not significantly expanding after its stars formed. Tidal shocking \citep[e.g.][]{2006MNRAS.366..429R} and/or collisionless heating due to SN feedback \citep[e.g.][]{2005MNRAS.356..107R,2013MNRAS.429.3068T} could both cause the stellar distribution to expand. For Sextans, which had its last burst long ago, this could be a potential worry; for Carina, which had its last burst very recently, the effect should be small (see Fig. \ref{fig:sfhs}). In \S\ref{sec:discussion}, we show that additional constraints from pulsar dispersion and X-ray emission measurements give 
an independent upper bound that is consistent or stronger than that derived from pressure confinement. This suggests that our assumption that  $r_\mathrm{gas}|_\mathrm{min} \sim r_\mathrm{SF}$ is sound.

In practice, we must solve equation (\ref{eqn:rgas}) {\it iteratively} since we do not know $\rho_{\rm cor}$, yet we require $r_\mathrm{gas}$ to calculate $\rho_{\rm cor}$. We describe this iterative calculation in \S\ref{sec:initialConditions} where a more realistic gas distribution, derived from the reconstruction of the SFH of the dwarf galaxy, is also used.

\subsection{Tidal stripping and shocking}\label{sec:tides}

In addition to ram-pressure stripping, dwarf galaxies will also experience tidal stripping and shocking. Tidal stripping becomes important roughly when the dynamical density of the dwarf matches the dynamical density of the host galaxy. As for ram-pressure stripping, this is most effective at pericentre \citep[e.g.][]{2006MNRAS.366..429R}:
\begin{equation} 
r_{\rm t} \sim \left[\frac{M_{\rm d}}{3M_{\rm h}}\right]^{1/3} r_{\rm p}\ , 
\end{equation} 
where $M_{\rm h}$ and $M_{\rm d}$ are the dynamical masses of the host galaxy and the dwarf, respectively, and $r_{\rm t}$ is the tidal stripping radius outside of which tidal stripping will become important. For typical MW dwarf galaxies like those we consider here, $r_{\rm p} \simgt 30$\,kpc \citep[e.g.][]{Lux+10}, $M_{\rm d} \simgt 3 \times 10^7$\,M$_\odot$ \citep[e.g.][and see Table \ref{tab:properties}]{Walker+09}, and $M_{\rm h}(<r_{\rm p}) \sim 2 \times 10^{11}$\,M$_\odot$ \citep[e.g.][]{2002ApJ...573..597K}. This gives $r_{\rm t} \simgt 1.1$\ kpc, which agrees well with the more careful analysis presented in \citet{2006MNRAS.367..387R}. 

Whether significant gas will be tidally stripped from the dwarf then depends on whether the dwarf ISM extends beyond the tidal stripping radius. Using equation (\ref{eqn:rgas}) and assuming a typical gas mass of $M_\mathrm{gas} \sim 10^6$\,M$_\odot$; a coronal density of $n_{\rm cor} \sim 2 \times 10^{-4} \cm$; and $T_\mathrm{gas} / T_\mathrm{cor} \sim 0.01$ gives $r_\mathrm{gas} \sim 0.9$\,kpc. Thus, $r_\mathrm{gas} < r_{\rm t}$ and we do not expect the gas in the dwarf to experience significant tidal stripping (see also a similar calculation in \citealt{BlitzRobishaw00}). \citet{2006MNRAS.367..387R} also estimate the likely effect of tidal shocking, finding that it is unimportant unless $r_p \simlt 20$\,kpc which is unlikely for the dwarfs we study here \citep{Lux+10}.

For the above reasons, we model only the ram pressure stripping of the dSphs in this work, deferring tides and/or other collisionless heating effects to future work.

\subsection{Adiabatic versus isothermal coronae}\label{sec:adiiso}
While it is likely that the MW has a hot corona of gas, it remains unclear what its thermodynamic state should be. Recent cosmological simulations produce a hot corona that is neither isothermal nor adiabatic \citep{2010MNRAS.407.1403C}, although these simulations are presently unable to make fully ab-initio predictions for disc galaxies in the real Universe \citep[e.g.][]{2008ASL.....1....7M}. For this reason, we consider here three cases of a fully isothermal, a fully adiabatic and an intermediate-state (so-called `cooling') corona. Assuming a polytropic equation of state $P = A \rho^\gamma$ for the gas, spherical symmetry, a background potential model for the MW $\Phi(r)$, and hydrostatic equilibrium, we may calculate the expected gas density profile $\rho$ by balancing pressure forces and gravity ($\nabla p = -\rho \nabla \Phi$; e.g. \bcite{Binney+09}) which gives: 
\begin{equation}
\rho = \left\{\begin{array}{lr}
\rho_0 \left[1-(\Phi - \Phi_0)\frac{\gamma - 1}{\gamma A}\right]^{\frac{1}{\gamma-1}} & \gamma \neq 1 \\
\rho_0 \exp\left(-\frac{\Phi-\Phi_0}{A}\right) & \gamma = 1 \end{array}\right.\ ,
\label{eqn:rhodist} 
\end{equation} 
where $\rho_{\rm 0}$ and $\Phi_{\rm 0}$ are, respectively, the density and potential at
the reference radius $r_{\rm 0}$. For isothermal haloes, $\gamma = 1$ and we may
write $P = A \rho \propto \rho T$ and therefore $T = T_0 = \mathrm{const.}$,
as expected. For adiabatic haloes, $\gamma = 5/3$. Thus, we consider models in
the range $1 \leqslant \gamma \leqslant 5/3$. Note that the potential $\Phi(r)$, $\rho_0$,
$\gamma$ and $A$ are all effectively free parameters in this model which must be
matched to the MW.

Throughout the paper, we will assume a truncated flat (TF) potential model for
the MW \citep{Wilkinson&Evans99}: 
\begin{equation}
\Phi(r)=- \dfrac{GM}{a}\mathrm{ln}\Bigl(\dfrac{\sqrt{r^2+a^2}+a}{r}\Bigr)\ ,
\label{TF}
\end{equation}
with $a=170$ kpc and $M=1.9\cdot10^{12}$ M$_\odot$. This was one of two profiles
used by \citet{Lux+10} to determine the orbits the MW dwarfs, and for
consistency we use the same potential in all our calculations.
\citet{Lux+10} found that within current observational uncertainties,
the choice of potential does not significantly affect the orbit
determination. In fact, for highly eccentric orbits, only the potential at
pericentre $\Phi(r_{\rm p})$ is relevant for the purpose of this
work\footnotemark; at $30\lsim r_{\rm p} \lsim 100$\,kpc this is reasonably well constrained for the MW \citep[e.g.][]{GD2}.

\footnotetext{To see why that is the case, notice that in the limit $r_{\rm a} \gg r_{\rm p}$,
and assuming spherical symmetry, the pericentre of the dwarf orbit is completely
determined by its specific angular momentum \citep[e.g.][]{2006MNRAS.366..429R}:
\begin{equation}
J^2 \simeq -2r_{\rm p}^2 \Phi(r_{\rm p})\ ,
\label{eqn:Jlimit}
\end{equation}
while the velocity at pericentre is then simply $v_{\rm p} = J / r_{\rm p}$. The dwarf's
specific angular momentum $J$ simply follows from its current distance from the
centre of the MW and its tangential velocity $J = d v_{\rm t}$ that comes from
a mixture of its doppler velocity and proper motion (depending on its
orientation on the sky). Thus, for eccentric orbits, $r_{\rm p}$ follows
observationally from a measurement of $J$, and a model assumption about
$\Phi(r_{\rm p})$.}

If we probe only one -- or several very similar -- $r_{\rm p}$ across several dwarfs,
then equation (\ref{eqn:rhodist}) is only required to {\it extrapolate} our
results to larger and smaller radii. We must assume some value for the coronal
temperature $T_{\rm cor}(r_{\rm p})$. But we may then after-the-fact assume an adiabatic or
isothermal corona and explore what this means e.g. for the missing baryon
fraction in the MW (see \S \ref{sec:missingBaryons}). If, however,
we have data at multiple $r_{\rm p}$ of wide separation then we must specify a model
up-front since the temperature $T_{\rm cor}$ (required to calculate $r_\mathrm{gas}$;
see equation (\ref{eqn:rgas})) is in general a function of radius through equation (\ref{eqn:rhodist}): 
we must perform a joint analysis of all dwarfs simultaneously.
For the moment, we restrict our analysis to two dwarfs (Sextans and Carina)
with very similar $r_{\rm p}$ within their uncertainties (see Table
\ref{tab:properties}). 
We model each dwarf separately using this independent analysis as a consistency check of our methodology and its assumptions.
In the future, it would be interesting to analyse the only dwarf (Fornax) with
pericentre radius significantly different from Carina and Sextans ($r_{\rm p}=110\pm20 \kpc$).
However, Fornax is 30 times more luminous than our two dwarfs and to obtain reliable results from the 
simulations will be much more challenging since it would require about $10^2$ times more grid points than our current simulations (and possibly 3D simulations).

\section{Method}\label{sec:method}
We use the code {\sc ECHO++} \citep{Marinacci+10,Marinacci+11}, an Eulerian
fixed-grid code based on \citet{DelZanna+07},
to run a series of high-resolution, two-dimensional hydrodynamical simulations
of dwarf galaxies moving through a hot rarefied medium, representative of the
Galactic corona. The simulations were performed over a Cartesian
grid with open boundaries. The dwarf galaxy is located on the $y = 0$ axis and
embedded in a hot medium, which moves along the $x$-axis with a speed that varies
with time, allowing us to model the motion of the dwarf along its orbit.
The simulations include both radiative cooling and SN
feedback. In the following subsections we describe the initial conditions.

\subsection{DM and coronal gas}\label{sec:darkMatter}
We set up a dwarf galaxy as a spherical distribution of cold isothermal gas
($T_{\rm gas}=1 \times 10^4 \K$) in hydrostatic equilibrium in a fixed potential. Given
that dSphs have mass-to-light ratios typically above 10 \citep[e.g.,][]{Battaglia+08},
we assume that the potential
is totally dominated by DM, and we neglect both the stellar mass
and the self gravity of the gas. The gravitational force, which determines the
initial cold gas distribution profile and at the same time counteracts the
ram pressure, is computed by using the NFW profiles taken from \citet{Walker+09}
(see Table \ref{tab:properties}). For each dwarf, the spherical DM halo is
located at the centre of the computational box.
Note that the DM parameters used in this work
refer to present-day observations. Since tidal stripping can remove DM from these haloes,
we are potentially underestimating the gravitational restoring force which acted against 
the ram-pressure at the time of the last stripping event. 
As a consequence the coronal value recovered with Sextans might in principle be higher than
the value obtained, while Carina should not be affected given that the last 
stripping event occurred very recently.

The cold medium of the dwarf is in pressure equilibrium with an external hot
medium, which represents the Galactic corona. 
This hot medium fills the whole computational box and it is assumed to
have constant density, temperature and metallicity; it moves along the x-axis
with a velocity that depends on the orbital path of the dwarf (see \S
\ref{sec:orbits}). The metallicity of the corona is fixed to $0.1\ Z_{\odot}$
in agreement with the recent observational determinations for NGC 891 \citep{Hodges-Kluck&Bregman13},
while our default corona temperature is $T_{\rm cor} = 1.8 \times 10^6 \K$
\citep{Fukugita&Peebles06}. Different temperatures for the coronal gas are 
investigated in \S\ref{sec:temperatures}.

The last parameter to set is the number density of the coronal gas $\ncor$, which is the goal of our investigation. In the following, with
$\ncor$ we refer to the total number density, which for a
completely ionized medium is the sum of the number density of ions $n_{\rm i}$ and
of electrons $n_{\rm e}$. We assume an abundance of helium of 26.4\% from
big bang nucleosynthesis considerations, which makes the electron density
$n_{\rm e}\simeq 1.1 \, n_{\rm i}$. The coronal density is then found
iteratively, by running several simulations and finding the value that produces
the complete stripping of cold gas from the dwarf at the end of the run. Note
that $\ncor$ also sets the pressure of the external medium, which in turn
determines the radius at which the pressure equilibrium is reached. We return to
this point in \S \ref{sec:initialConditions}.

\subsection{Orbits}\label{sec:orbits}
One of the basic parameters that have to be set in our simulations
is the relative velocity between the dwarf and the surrounding medium, 
which requires the knowledge of the orbital path of the satellite galaxy
in the potential of the MW. For this purpose, we use the
reconstruction of the dwarf orbits derived by
\citet{Lux+10}. These authors provide a set of 1000 possible orbits for each
dwarf given the potential of the MW and the, unfortunately poorly constrained,
proper motions of the dwarfs. They considered two Galactic potentials: the
TF model \citep[][and equation \ref{TF}]{Wilkinson&Evans99} and the
\citet{Law+05} model. In this work we only use the former. As discussed in
\S \ref{sec:adiiso}, we are not very sensitive to the choice of potential
models: uncertainties in the orbit coming from proper motion errors and other
model systematics will dominate our error budget. When {\it extrapolating} our
results to larger radii, however, the choice of the potential model and
the assumed thermodynamic state of the hot coronal gas become important.
We discuss this further in \S\ref{sec:missingBaryons}. 

The families of orbits for each dSph are classified in terms of the pericentric
radius ($r_{\rm p}$) and the velocity at pericentre ($v_{\rm p}$). We select
only the orbits having pericentric passages compatible with our estimate of
look-back time of the last burst of star formation $t_{\rm lb}$ (see \S
\ref{sec:initialConditions}). In practice, given $t_{\rm lb}$ and the
width of the last SFR temporal bin, we accept only the orbits which have a
pericentric passage within this bin.
\begin{figure}
\includegraphics[width=0.5\textwidth]{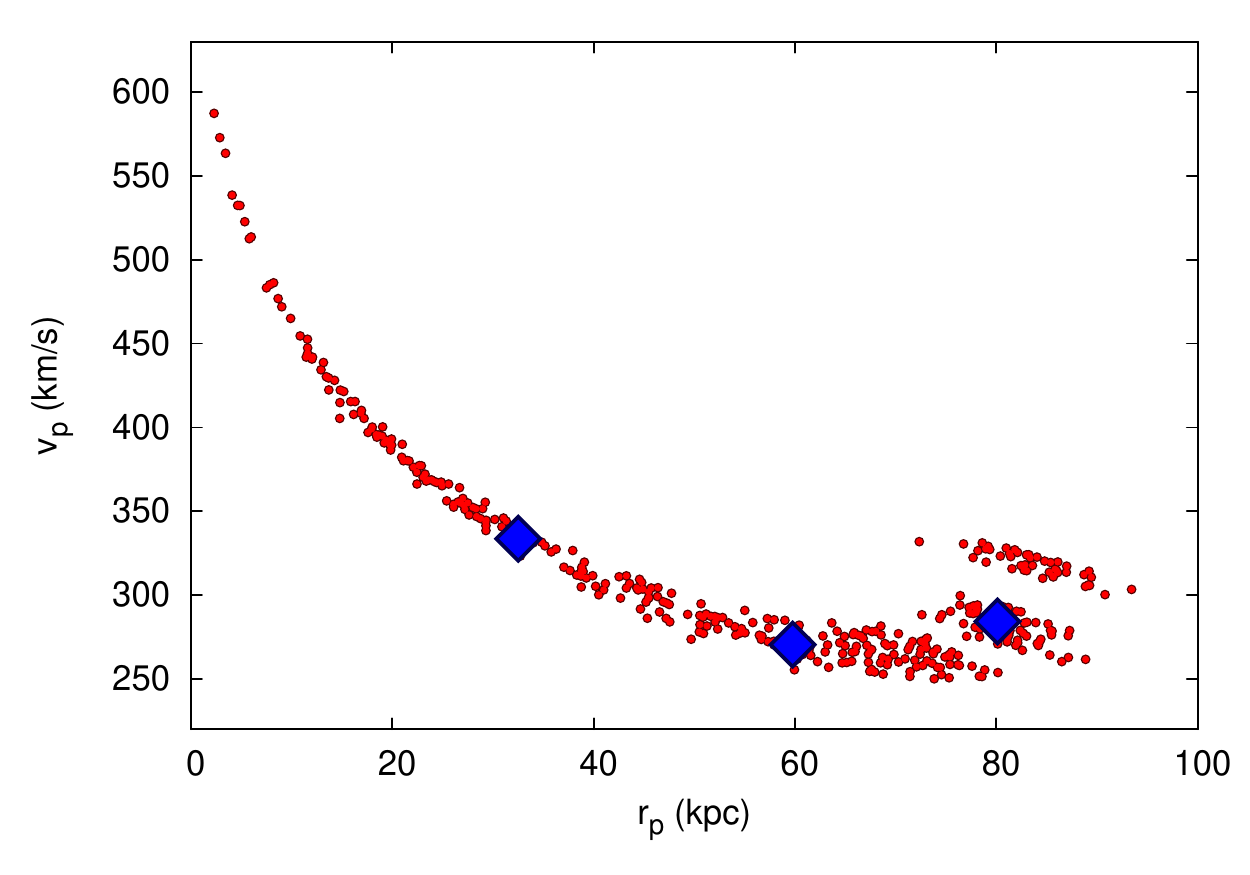}
\caption{Pericentric radii and velocities for the orbits of the Sextans dSph
compatible with a pericentric passage at the stripping time ($t_{\rm lb}$)
determined from the SFH (see the text). The large (blue) diamonds show the three
representative orbits (median, first and third quartiles in $r_{\rm p}$ and in
$v_{\rm p}$ within $\pm 3$ kpc from the selected value of $r_{\rm p}$) chosen
for our simulations.} 
\label{fig:orbits}
\end{figure}
Fig.\ \ref{fig:orbits} shows the distribution of these orbits in
the ($r_{\rm p}$, $v_{\rm p}$) space. Given the non-triviality of this
distribution we decide to focus on three representative orbits. The median orbit
is given by the median value of the pericentric radius $\bar{r}_{\rm p}$ and the
median of $\bar{v}_{\rm p}$ in the range $\pm 3\,\kpc$ around $\bar{r}_{\rm p}$.
For Sextans, we obtain $\bar{r}_{\rm p}=59.8\,\kpc$ and $\bar{v}_{\rm p}=270.4
\kms$, in agreement with the values obtained by \citet{Lux+10} for the last
pericentric passage. We then select two more orbits at the first and third
quartiles of the distribution of $r_{\rm p}$ and their corresponding values of
$v_{\rm p}$. The selected orbits are indicated by the large diamonds 
in Fig.\ \ref{fig:orbits}.

Thus, as far as the orbits are concerned, we perform three distinct sets of
hydrodynamical simulations. The parameters of the three representative orbits
for Sextans are reported in Table \ref{tab:simulations}. We show in \S\ref{sec:densitySex} that the results for Sextans' orbits are remarkably
consistent with each other despite the large difference in their input
parameters. This is an encouraging test of our model assumptions and
systematics. Given these results, we consider only the median orbit for Carina.

The typical orbital periods of our two dwarfs are between 1 and 4\,$\Gyr$. However,
the stripping process is much more efficient at and near the orbital pericentre
and we can save computational time by simulating only that part of the orbit.
To have an idea of the variation of the stripping efficiency within an orbit one
can make use of equation (\ref{eqn:crudestripform}) to obtain:
\begin{equation}
\label{eq:baseefficiency}
\varepsilon_\mathrm{strip}(r)=\dfrac{v(r)^2}{v_{\rm p}^2} \dfrac{\ncor(r)}{\ncorp}\ ,
\end{equation}
where $v(r)$ and $r(t)$ are the position and velocity of the dwarf along its
orbit at time $t$; $\ncor(r)$ is the coronal density at $r$; and 
$\ncorp$ is the coronal density at pericentre.
For our two dwarfs, the efficiency calculated from equation (\ref{eq:baseefficiency})
changes by a factor of $\sim$10 from pericentre to apocentre.
After performing a series of simulations progressively enlarging the computational 
time up to the full length of the orbit we find that including in the calculation
regions where the efficiency has dropped below 50\% from the pericentre does not
result in any appreciable difference in the derived coronal density.
Thus, we focus on the part of the orbit with efficiency above 50\%, which leads to
the integration times reported in Tables \ref{tab:simsetup} and \ref{tab:simulations}.
The $x$-component of the
velocity of the hot gas is set according to the relative velocity $v(r(t))$,
which in turn depends on the selected orbit.
For simplicity, we keep the value of the coronal density constant in our
simulation. In this way we derive an {\it average} value of the coronal
density over the orbit segment around the pericentre. 
Finally, we vary $\ncor$ until we find the value
that produces a complete removal of gas from the dwarf galaxy: $\ncormin$.

\subsection{Initial gas distribution}\label{sec:initialConditions}
\begin{table*}
\centering
\begin{tabular}{lcccccccc}
\hline
    dSph    & $r_{\rm SF}$ &$t_{\rm lb}$ & SFR & $\Sigma_{\rm SFR}$ & $\bar{n}_{\rm gas}$ & $n_{\rm 0, gas}$ & $r_{\rm gas}$ & $M_{\rm gas}$\\
            & (kpc)        & (Gyr)   & ($\rm\moyr$) & ($\rm\moyrkpc$) & ($\cm$) & ($\cm$) & (kpc) & ($\rm\mo$)\\
\hline
Sextans & 0.5  & 7 & $4.6\pm2.2\times10^{-5}$ & $5.9\times10^{-5}$ & 0.09 & 0.27 & 0.98 & $7\times10^6$ \\ 
Carina  & 0.28 & 0.5 & $4.6\pm1.3\times10^{-6}$ & $1.9\times10^{-5}$ & 0.14 & 0.4  & 0.4  & $6.3\times10^5$ \\ 
\hline
\end{tabular}
\caption{Star formation properties and derived cold gas content of our two
dSphs at the time of the last ram-pressure stripping event. The SFRs and gas
density distributions are used as initial conditions for our hydrodynamical
simulations. From left to right, the columns show the radius at which the SFR has been extrapolated;
time to the last star formation burst; star formation rate at $t_{\rm lb}$; SFR surface density at $t_{\rm lb}$ $\Sigma_{\rm SFR}=\mathrm{SFR}/(\pi r_\mathrm{SF}^2)$; initial mean cold gas density; initial central
gas density; computed radius of the cold gas distribution; computed initial cold
gas mass within $r_{\rm gas}$.}
\label{tab:initialConditions}
\end{table*}

\begin{figure}
\centering
\includegraphics[width=0.5\textwidth]{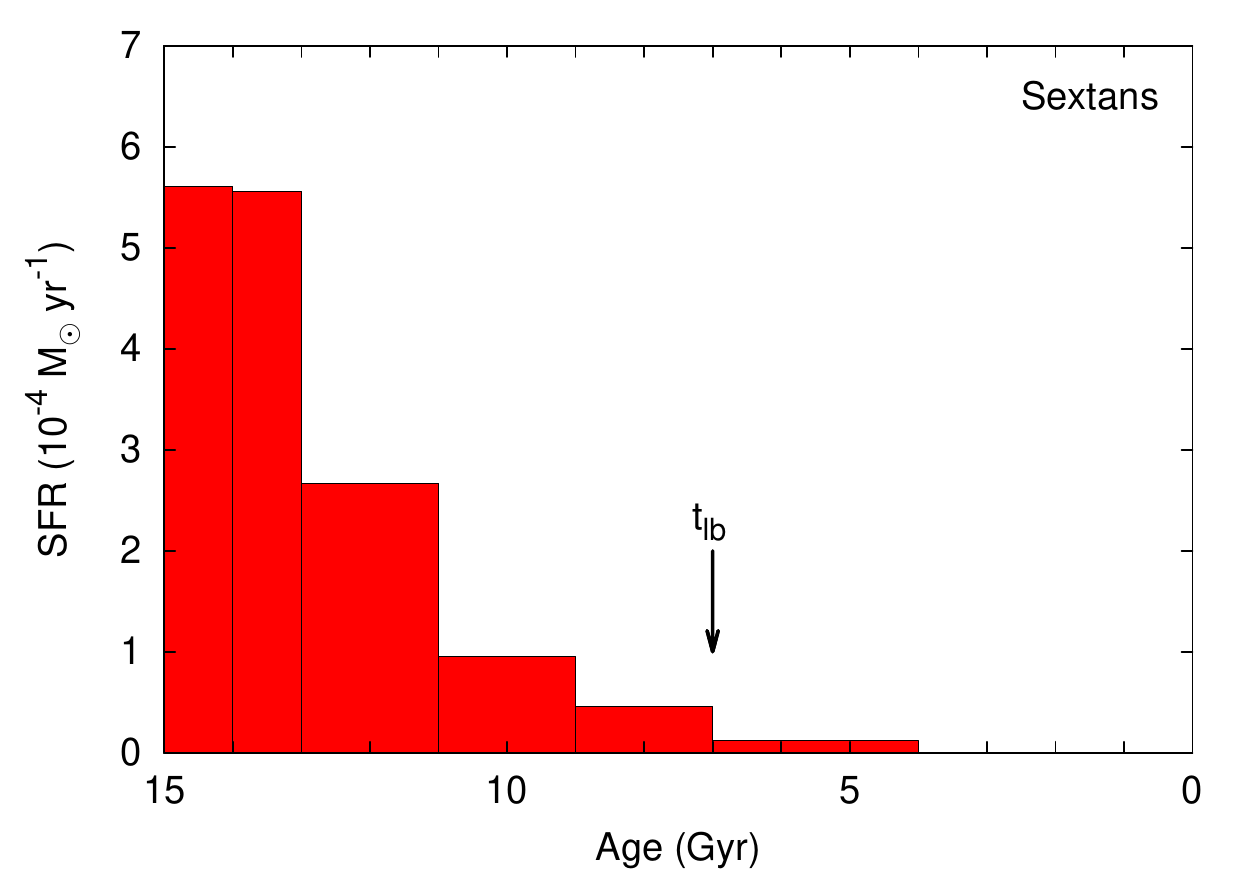}
 \includegraphics[width=0.5\textwidth]{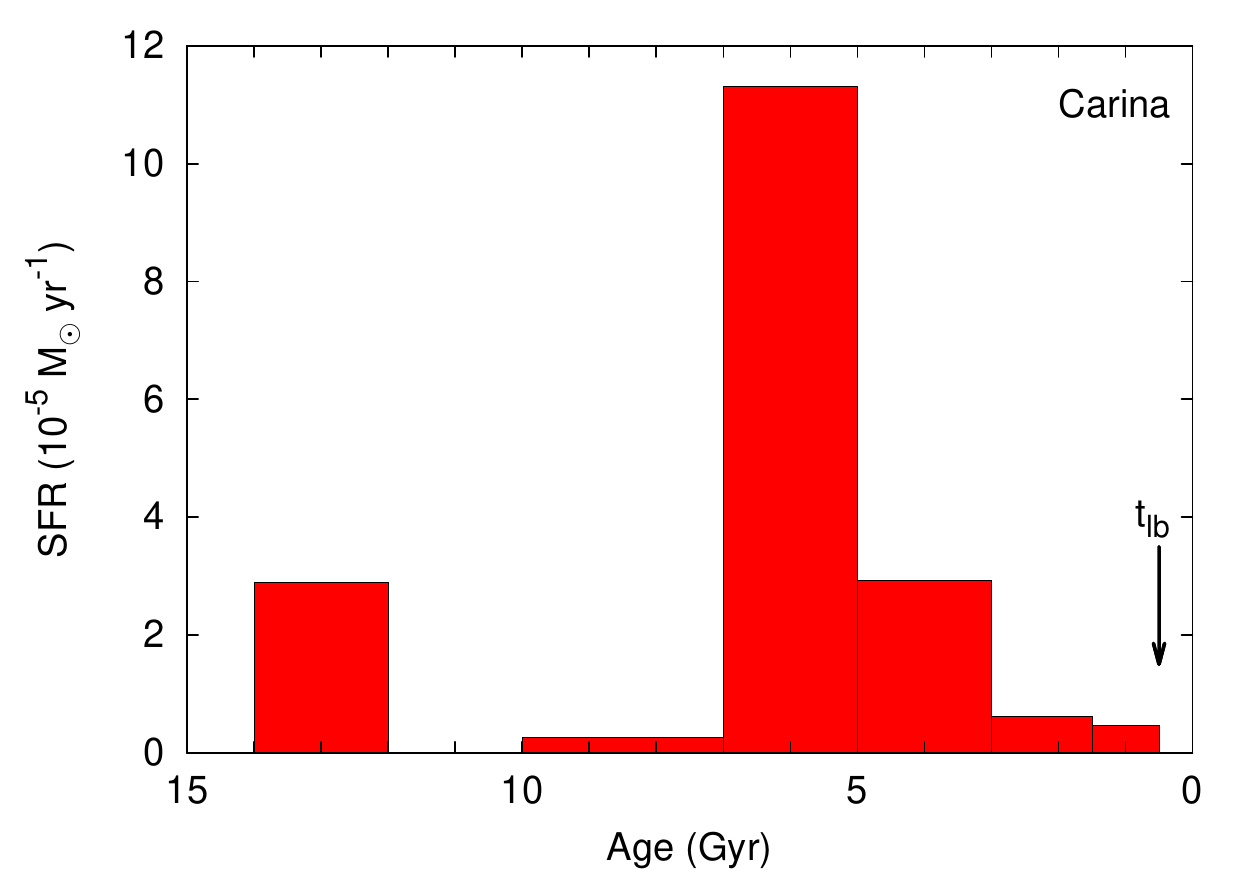}
 \caption{SFH of Sextans from \citet{Lee+09} and of Carina from \citet{Rizzi+03}. 
The arrow indicates the look-back time of last burst of star formation.
In our scheme, this corresponds to the last stripping event.}
 \label{fig:sfhs}
\end{figure}

In our simulations, the ISM of the dwarf galaxies is composed by isothermal
($T=10^4 \K$) gas that is in hydrostatic equilibrium with the DM potential and
has a subsolar metallicity taken from the literature (see Table \ref{tab:simsetup}).
Note that the metallicity of the coronal gas is always set to 0.1\,$Z_{\odot}$ (see \S \ref{sec:darkMatter}).
The radius at which the cold gas distribution is truncated corresponds to the radius
where its pressure is equal to the pressure of the coronal gas.
The latter depends on the coronal temperature, for which we explore different values,
$T_{\rm cor} = 1,\ 1.8,\ 3 \times 10^6 \K$ (see again \S \ref{sec:darkMatter} and \ref{sec:temperatures}).
Since the gravitational potential of the DM halo is fixed, the gas density distribution in
the dwarf is fully determined once we set the central density. We estimate this
central density using information contained in the SFH,
as described below.

We derived the look-back time of the last burst of star formation
($t_{\rm lb}$) as the time when the estimated value of the SFR is consistent
with zero within the given uncertainties.
At that time, we assume that the dwarf has a negligible amount of gas left, 
i.e.\ we consider the gas stripping process as completed. 
In Table \ref{tab:properties}, we report the times
of the last stripping event for
the dSphs. We refer to this as the {\it last} stripping event because it is
likely that dSphs have suffered gas stripping also at earlier times. 
Considering only the last event has a number of
advantages: (i) it saves computational time; (ii) it allows us to probe the
corona at the closest possible look-back time; and (iii) most importantly, 
it allows for the best possible reconstruction of the orbital paths
(see \S \ref{sec:orbits}).

The SFHs of Sextans and Carina, taken from \citet{Lee+09,Rizzi+03}, are 
shown in Fig.\ \ref{fig:sfhs}. The look-back times of the last starburst (i.e., the last stripping event) are 
$t_{\rm lb}\sim 7$ and $0.5 \Gyr$, respectively. From the SFHs we can then extract the SFRs at the time prior to this
event. These values are reported in Table \ref{tab:initialConditions}.

There are two key uncertainties related to our reconstruction of the SFR at a given look-back time:

\begin{enumerate} 
\item The time resolution of the SFH makes the $t_{\rm lb}$ uncertain by 
about $0.5-1 \Gyr$. This is a small error compared to other uncertainties.
\item The presence of `blue straggler stars' may contaminate the SFH, masquerading as recent star formation.
\citet{Lee+09} explicitly consider this, publishing an alternate `corrected' SFH for Sextans. The corrected
SFH has no star formation at $t > t_{\rm lb}$, and a small reduction in star formation at $t_{\rm lb}$. 
We preferred to use the uncorrected SFH shown in Fig.\ \ref{fig:sfhs} because it is consistent
with the one used for Carina (where the correction has not been applied). However, we explicitly tested for
the effect of blue straggler contamination on our results by running some additional simulations using the
corrected SFH of \citet{Lee+09}. We found that the final value of the coronal density does not vary appreciably
within our quoted uncertainties.
\end{enumerate}
Once we know the SFR before stripping, we use a revised version of the
K-S relation to estimate the gas
density at that time. The standard K-S relation connects
the (molecular and atomic) hydrogen surface density, $\Sigma_\mathrm{\hi}$ and
$\Sigma_\mathrm{H_2}$, and the SFR surface density, $\Sigma_\mathrm{SFR}$,
with a power-law (slope 1.4). It is valid for disc galaxies and starburst galaxies
\citep[e.g,][]{Kennicutt98b}. It is well known that this relation steepens
considerably for column densities below $\sim 10 \rm\mopc$
\citep[e.g.][]{Leroy+08}. While the $\Sigma_\mathrm{SFR}$ seems to correlate
very well with the molecular gas surface density \citep{Bigiel+11}, the relation breaks down at
low densities likely due to the transition from a molecular-dominated to an
atomic-dominated ISM \citep{Krumholz+12}. Due to the low values of the SFRs of
our dwarfs (see Table \ref{tab:initialConditions}), the expected
$\Sigma_\mathrm{\hi+H_2}$ falls below the limit and the dwarfs' ISM is dominated
by \hi, as confirmed by observations (see Table \ref{tab:dIrrs} and references
therein). In this paper, we do not distinguish between different gas phases in
the ISM as in the simulations the cooling is 
truncated at $10^4$\,K. This is an acceptable
approximation since our star formation and feedback prescriptions are purely
empirical and based on the observed SFR. 

\begin{figure}
\centering
\includegraphics[width=0.5\textwidth]{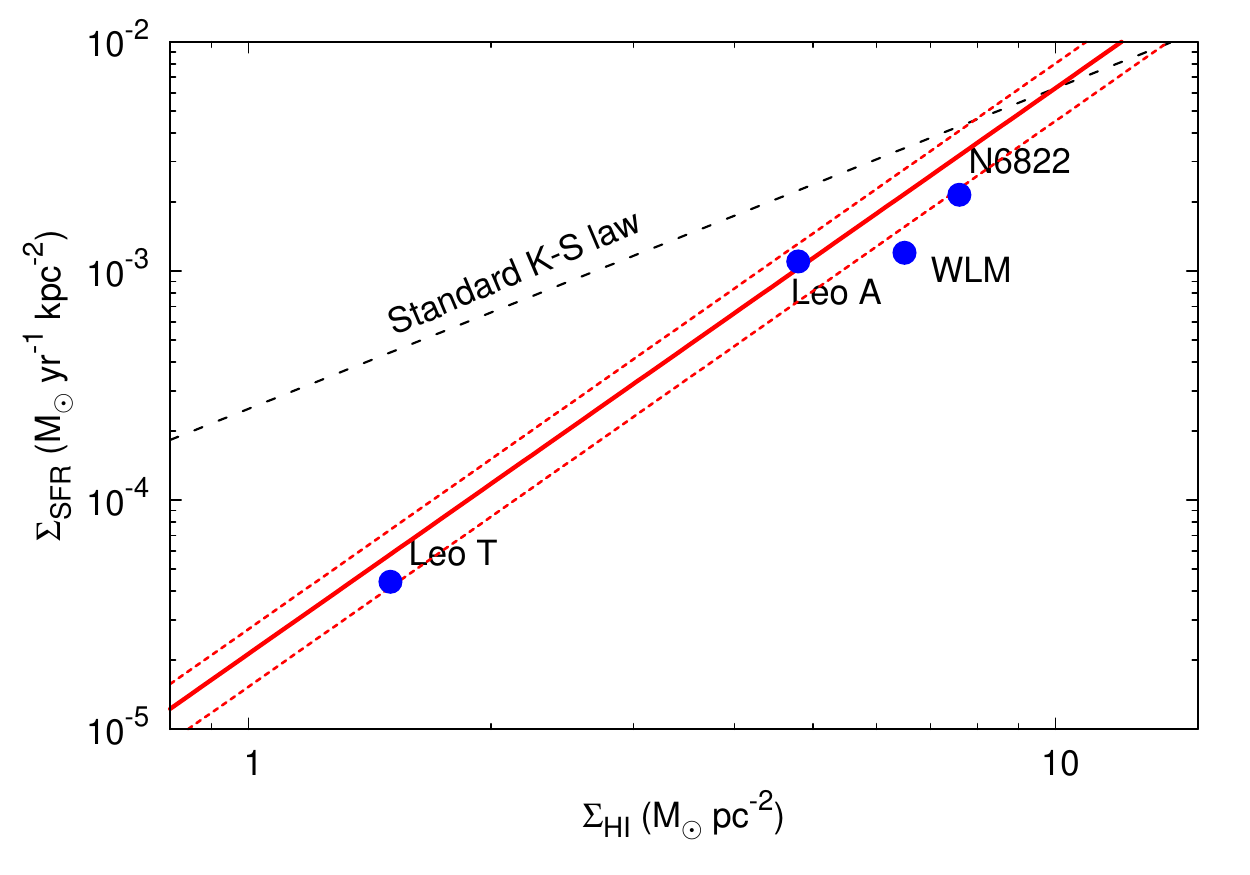}
\caption{K-S relation for dwarf galaxies as described by equation (\ref{eq:SFlaw}). 
The points show the location of four Local Group dIrrs
(see Table \ref{tab:dIrrs}).} 
\label{fig:SFlaw}
\end{figure}
\begin{table}
\centering
\begin{tabular}{lcccc}
\hline
Galaxy & $\Sigma_\mathrm{SFR}$ & $\Sigma_\mathrm{HI}$ & $\Sigma_\mathrm{H_2}$ &Ref.\\
       & ($\moyrkpc$)   &\multicolumn{2}{l}{\ \ \ ($\rm\mopc$)}  & \\
\hline   
NGC 6822 & $2.15\times10^{-3}$ & 7.6       & 1.1        &(1,2,3) \\
WLM      & $1.2\times10^{-3}$  & 6.5       & Negligible &(4,5,6) \\
Leo A    & $1.1\times10^{-3}$  & 4.8       & Missing    &(7,8) \\
Leo T    & $4.4\times10^{-5}$  & 1.5       & Missing    &(9,10) \\
\hline
\end{tabular}
\caption{SFR densities and gas densities for four dIrrs of the Local Group. CO
is not detected in Wolf-Lundmark-Melotte (WLM), and there is only an upper limit, while for Leo A and Leo T
such studies are missing in the literature. References to the SFR, \hi\ and CO
studies (when applicable): (1) \citet{Efremova+11}, (2) \citet{deBlok&Walter06},
(3) \citet{Israel97}, (4) \citet{Dolphin00}, (5) \citet{Kepley+07}, (6)
\citet{Taylor&Klein01}, (7) \citet{Cole+07}, (8) \citet{Young&Lo96}, (9)
\citet{deJong+08}, (10) \citet{Ryan-Weber+08}.}
\label{tab:dIrrs}
\end{table}

To date, there is no consensus on how to extend the K-S relation to surface densities $< 10 \rm\mopc$.
Some authors have however studied the location of dwarf galaxies in
the ($\Sigma_\mathrm{SFR}$, $\Sigma_\mathrm{\hi}$) plane.
\citet{Bigiel+10} studied five dwarf galaxies and found a relation
$\Sigma_\mathrm{SFR}\propto\Sigma_\mathrm{\hi}^{1.7}$.
Using a larger sample of 23 very faint dwarf galaxies, \citet{Roychowdhury+09}
found that these systems depart systematically from the
standard K-S relation but they, quite remarkably, follow the Kennicutt relation for disc galaxies only
\citep[excluding starburst galaxies;][]{Kennicutt98a}.
This relation can be written as follows: 
\begin{equation} 
\label{eq:SFlaw}
\Sigma_\mathrm{SFR}=(2.13\pm0.6)\cdot10^{-5}\ \Sigma_\mathrm{gas}^{2.47}\ .
\end{equation}
Note that $\Sigma_\mathrm{SFR}$ is given in $\rm\moyrkpc$ and $\Sigma_\mathrm{gas}$
in $\rm\mopc$. 
In the following we adopt equation (\ref{eq:SFlaw}),
where the normalization factor and the associated errors (roughly $1\sigma$) have been
taken from the standard K-S relation using the normalization of \citet{Roychowdhury+09}.

To make sure that equation (\ref{eq:SFlaw}) is suitable
for our purposes, we check that it holds for galaxies in the Local Group. We
consider four dIrrs that span a large range of gas and SFR surface
densities. For each of them, we calculate $\Sigma_\mathrm{SFR}$ knowing the value of
the SFR and the area of the galaxy from which it has been derived.
We then estimate the surface densities of \hi\ and molecular
gas (when present) averaged over the same area. The obtained values are listed
in Table \ref{tab:dIrrs}. As expected, the molecular phase plays a minor role
and can safely be neglected. In Fig.\ \ref{fig:SFlaw} we show the obtained values
of $\Sigma_\mathrm{SFR}$ and $\Sigma_\mathrm{HI}$ (solid circles),
as well as the relation from equation (\ref{eq:SFlaw}). The
agreement is remarkably good for all the dIrrs,
the dashed lines show the $1\sigma$ error.
Note that the standard K-S relation (dashed line) would clearly overestimate
$\Sigma_\mathrm{SFR}$ at these gas surface densities by up to an order of magnitude.

Using $\Sigma_\mathrm{SFR}$ reported in Table \ref{tab:initialConditions}, we
estimate the average gas volumetric density (assuming spherical symmetry)
for the two dSphs by inverting equation (\ref{eq:SFlaw}): 
\begin{equation} 
\label{eq:rhoGas}
\overline{\rho}_{\rm gas}(<r_\mathrm{SF})=\dfrac{3}{4r_\mathrm{SF}}\Biggl(\dfrac{\Sigma_\mathrm{SFR}(<r_\mathrm{SF})}{2.13\times 10^{-5}}\Biggr)^{\frac{1}{2.47}}\ ,
\end{equation}
where $r_\mathrm{SF}$ is the radius within which the SFH has been derived\footnote{It
is worth mentioning that equation (\ref{eq:rhoGas}) derives from a slightly different definition of $\Sigma_\mathrm{gas}$.
This is due to the fact that our dwarfs have to be considered spheroidals, while equation (\ref{eq:SFlaw}) formally holds only for discs.}.
The gas density profile is then rescaled to match this average density within
$r_{\rm SF}$. This allows us to determine the central density $n_{0,\rm gas}$ and
the total gaseous mass of the dwarf within $r_{\rm gas}$, which is the radius at
which pressure equilibrium with the corona is reached. The densities
are then multiplied by a factor 1.36 to take into account the He fraction. All
these parameters are reported in Table \ref{tab:initialConditions}.

\subsection{Radiative cooling, star formation and feedback}\label{sec:feedback}
Radiative cooling is included in the code by taking the collisional ionization
equilibrium cooling function of \citet{Sutherland93}. The cooling
term is added explicitly to the energy equation of the gas and, for stability
reasons, the hydrodynamic time-step is reduced to 10\% of the minimum
cooling time in the computational domain. Metal cooling is taken into account
and the metallicity of the gas is treated as a passive scalar field advected by
the flow. The cooling rate is set to zero below $T_{\rm min} = 10^4$ K.

We include star formation in our hydrodynamical code by 
introducing a temperature cut, $T_\mathrm{cut}=4\times10^{4}
\K$. Only cells below this temperature are allowed to form stars. The amount of
gas converted into stars is computed from equation (\ref{eq:SFlaw}), where the gas density
is a function of time.
However, given that the star formation rates used for our simulated dwarfs are small (see Table
\ref{tab:initialConditions}), there is no significant depletion of gas. 
This is an important point as it shows that the removal of gas from 
Sextans and Carina can not be achieved by star formation alone. Rather, it requires additional
processes, i.e. a combination of SN feedback and gas stripping.

Concerning SN explosions, we assume that our SN bubbles start their expansion at the end of the adiabatic
(Sedov) phase and we only follow the subsequent radiative phase. In this phase,
the thermal energy is lost due to radiative cooling and adiabatic expansion, while the kinetic energy
is used partially for the expansion and partially it is transferred to the ambient medium at later times.
The explosion of a single SN is implemented by increasing the volumetric thermal energy
density by a factor $\frac{E_\mathrm{SN}}{V_\mathrm{Sedov}}$, where
$E_\mathrm{SN}=10^{51} \erg$ and $V_\mathrm{Sedov}=\frac{4}{3}\pi
r_\mathrm{Sedov}^3$ represents the initial spherical volume of the bubble,
with $r_\mathrm{Sedov}$ the radius of the injection region.
For every different gas profile, $r_\mathrm{Sedov}$ -- the SN
bubble radius at the end of the adiabatic phase -- is determined by running
very-high resolution simulations of a single SN exploding in the centre of the dwarf.
$r_\mathrm{Sedov}$ is then set to the value of the initial radius that produces
a match between the simulated evolution of the SN shock radius and the analytical
(two-dimensional) one for the radiative phase.  
We model a SN bubble at the explosion time with just four cells, since higher
numbers cause our simulations to be too demanding from a numerical point of view.
Thus, the resolution of a simulation is defined by the value of $r_\mathrm{Sedov}$
by simply equating the circular area of the SN bubble with the Cartesian one of four cells.  
We also adopt the overcooling correction method described in \citet{Anninos&Norman94}.
\begin{table*}
\centering
\begin{tabular}{lcccccccc}
\hline
    Run        & $L_{\rm box}$ & $\Delta\,x$ & $T_{\rm cor}$ & $n_{\rm 0, gas}$ & $Z$ & $\overline{v}_{\rm sat}$ & $\Delta\,r$  & $\Delta\,t$\\
                  & (kpc)         & (pc)       & (K)    & ($\cm$)                  &($Z_{\odot}$)      & (km/s)                   & (kpc)        & (Myr)\\
\hline
SextansMidMed     & 80  & 34 & $1.8 \times 10^6$ & 0.27 & 0.02 & 228 & 59.8--90.2  & 930\\
SextansLowMed     & 80  & 39 & $1.8 \times 10^6$ & 0.18 & 0.02 & 228 & 59.8--90.2  & 930\\
SextansMid1stQ    & 60  & 34 & $1.8 \times 10^6$ & 0.27 & 0.02 & 286 & 33.9--59.2  & 420\\
SextansLow1stQ    & 60  & 39 & $1.8 \times 10^6$ & 0.18 & 0.02 & 286 & 33.9--59.2  & 420\\
SextansMid3rdQ    & 100 & 34 & $1.8 \times 10^6$ & 0.27 & 0.02 & 246 & 80.4--131.5 & 1220\\
SextansLow3rdQ    & 100 & 39 & $1.8 \times 10^6$ & 0.18 & 0.02 & 246 & 80.4--131.5 & 1220\\
CarinaMidMed      & 80  & 31 & $1.8 \times 10^6$ & 0.4  & 0.01 & 251 & 51.2--81.8  & 740\\
CarinaLowMed      & 80  & 35 & $1.8 \times 10^6$ & 0.31 & 0.01 & 251 & 51.2--81.8  & 740\\
CarinaMidMed1e6K  & 80  & 31 & $1 \times 10^6$   & 0.4  & 0.01 & 251 & 51.2--81.8  & 740\\
CarinaMidMed3e6K  & 80  & 31 & $3 \times 10^6$   & 0.4  & 0.01 & 251 & 51.2--81.8  & 740\\
\hline
\end{tabular}
\caption{Parameters of the simulations. Each run is denoted by the
dwarf name, the initial density of the dwarf's ISM (see \S \ref{sec:initialConditions} and
\ref{sec:densitySex}), the pericentric distance of the orbit and the temperature of the corona 
(if different from the reference value $T_{\rm cor} = 1.8\times 10^6$ K). $L_{\rm box}$ is
the size of the computational domain in each direction, $\Delta x$ is the resolution, $T_{\rm cor}$ is the
coronal temperature, $n_{\rm 0, gas}$ is the initial central density of the
dwarf, $Z$ is the dwarf's gas metallicity, $\overline{v}_{\rm sat}$ is the dwarf
velocity averaged over the simulated distance range $\Delta\,r$ and $\Delta\,t$ is
the integration time corresponding to the part of the orbits with stripping
efficiency greater than 50\%.} 
\label{tab:simsetup}
\end{table*}

We compute the supernova rate (SNR) from the SFR using the initial mass function (IMF) $\Psi(M)$ chosen to retrieve the SFH
of our dwarf galaxies.
For Sextans and Carina, a Salpeter IMF \citep{Salpeter55} was assumed \citep[see][]{Rizzi+03, Lee+09}.
In this case $\mathrm{SNR}\simeq\frac{6\times 10^{-3}}{\rm M_\odot}\mathrm{SFR}$ $\frac{\mathrm{SN}}{\mathrm{yr}}$,
with the SFR expressed as $\rm M_\odot$ yr$^{-1}$.  
Applying for the SFR found in every cell with $T<T_{\rm cut}$ and multiplying the obtained SNR with the time-step,
we find the number of SN ``events'' occurring in each cell during a given time-step.
From this we can then generate random explosions across the dwarf galaxy.
Note that, since the SFR of the simulation is tied to the dwarf's gas, the SNR is
dependent on the amount of cold gas at that specific time-step,
assuming that the SNe form and explode instantaneously. 
Using this method the SNR in a simulation of a dwarf
in isolation (without the ``coronal wind'') is recovered within $\sim 10\%$ of the expected value
\footnote{SNe alone are inefficient in removing the gas (see also \S \ref{sec:SNe}). Thus, the SFR remains constant along the simulation
and the simulated SNR can be compared with the predicted one. The match between the computed and the expected value
(within 10\%) shows the viability of our implementation.}.

\subsection{Simulations setup}\label{sec:setup}
In Table \ref{tab:simsetup} we list the details of our main runs.
Different initial conditions for the dwarfs are computed by exploring the main model uncertainties:
the orbit reconstruction, the determination of the SFH and the star formation law (see \S \ref{sec:densitySex}).
Each set of runs for the two dwarfs has been simulated many times by changing the value of $\ncor$
(which sets the dwarf's gas truncation radius $r_{\rm gas}$ and initial mass
$M_{\rm gas}$ once the central gas density is fixed) until complete gas
stripping occurs at the end of the simulation.
We consider that a galaxy is devoid of gas when the mass of cold ($T<1 \times 10^5 \K$) gas bound
to the potential of the dwarf is $<5 \%$ of the initial mass.
The remaining small amount of cold gas can be easily stripped in the following
part of the orbit. Large sizes of the
computational box are needed to avoid boundary effects (such as reflected waves)
on the surface of the dwarf. The boundaries used are ``Wind'' in the x-direction
(``Inflow'' on the right side and ``Outflow'' on the left one) and ``Outflow''
in the y-direction. The velocity of the inflow is set according to the selected
orbits. $\Delta\,r$ (and the corresponding $\Delta\,t$) is determined by the orbit's choice,
and it represents the range of distances from the MW over which the
recovered coronal density has effectively been averaged. Such values have been
determined using equation (\ref{eq:baseefficiency}) with a stripping efficiency of 50\%
(see \S \ref{sec:orbits}).

\section{Results}\label{sec:results}
\begin{figure*}
\includegraphics[height=0.49\textwidth,angle=-90]{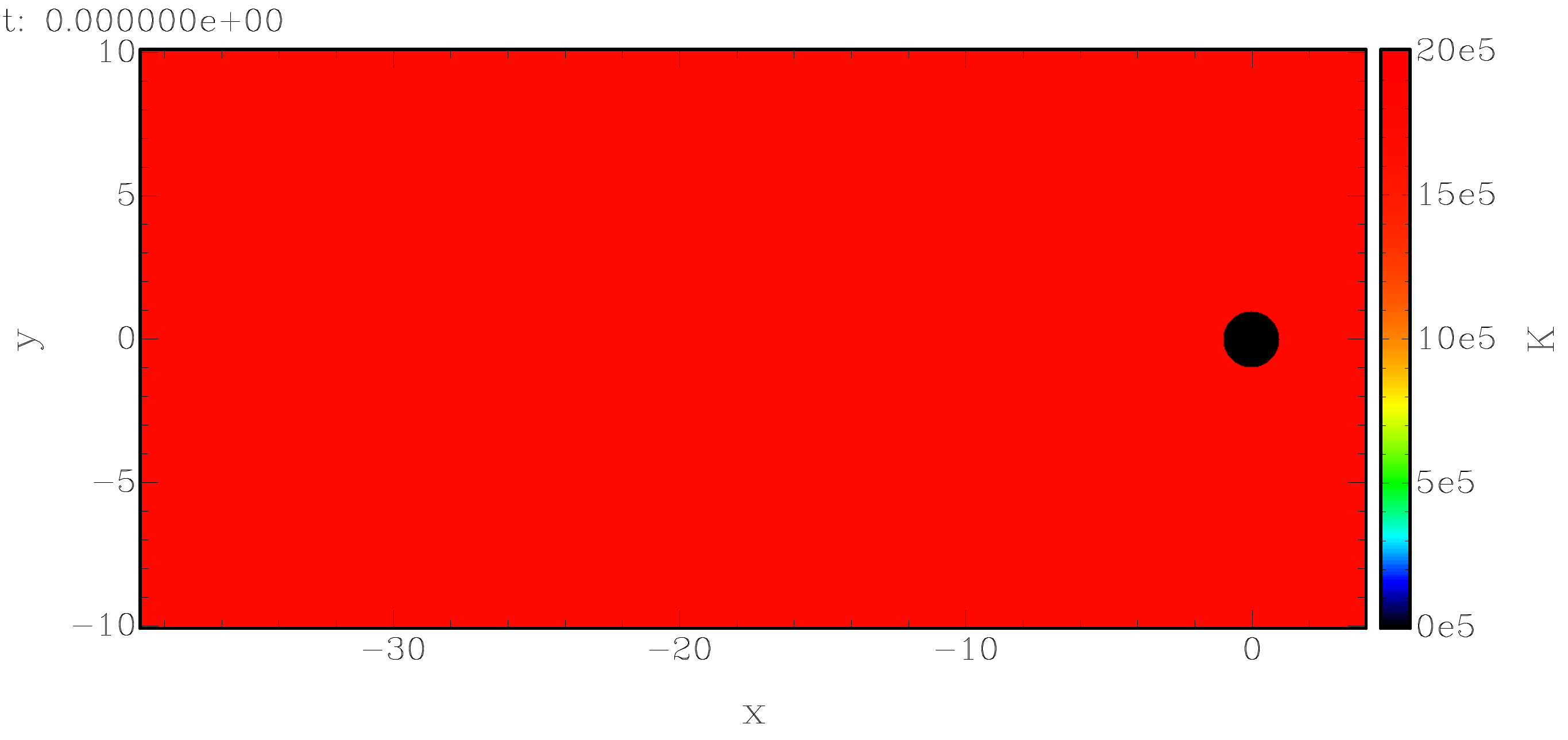}
\includegraphics[height=0.49\textwidth,angle=-90]{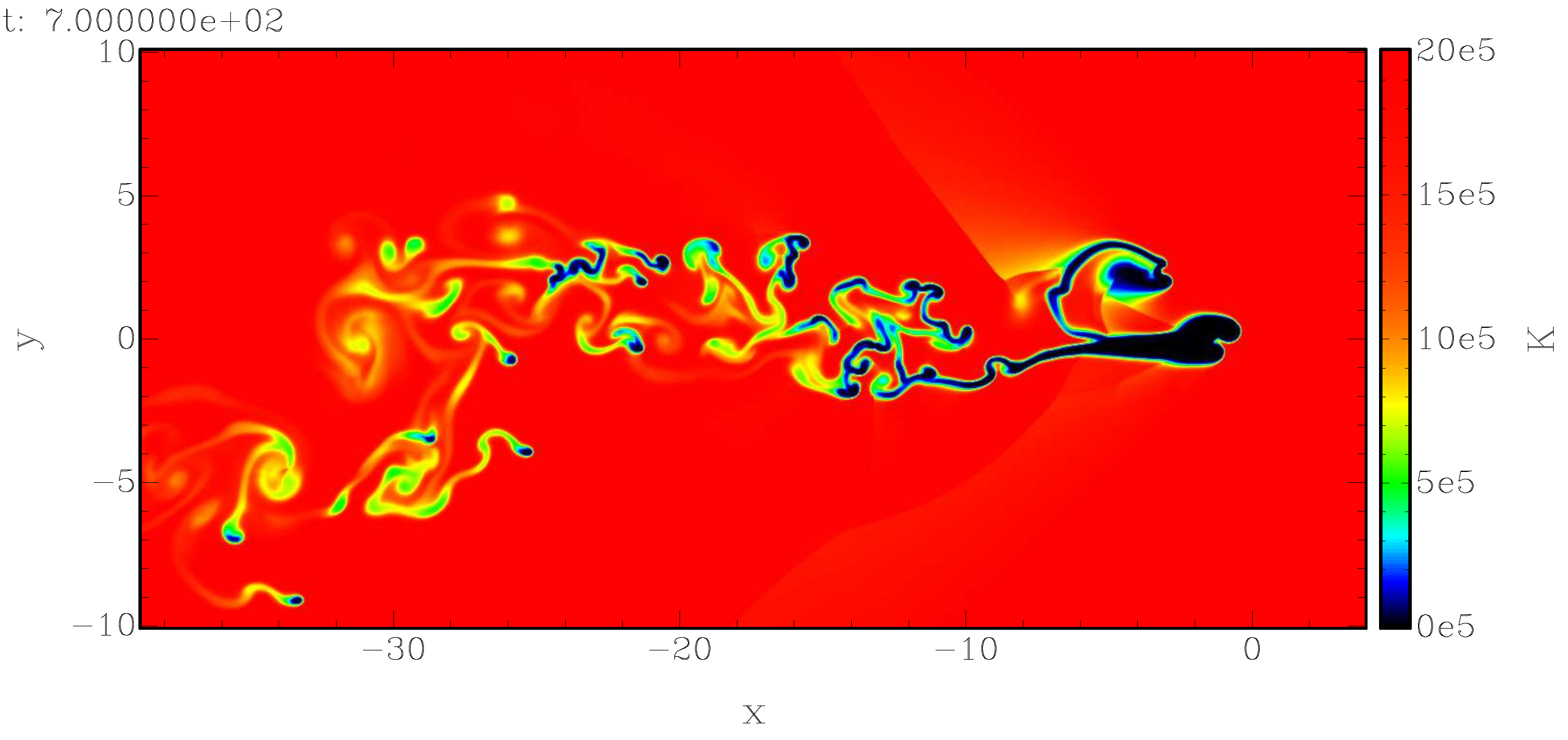}
\includegraphics[height=0.49\textwidth,angle=-90]{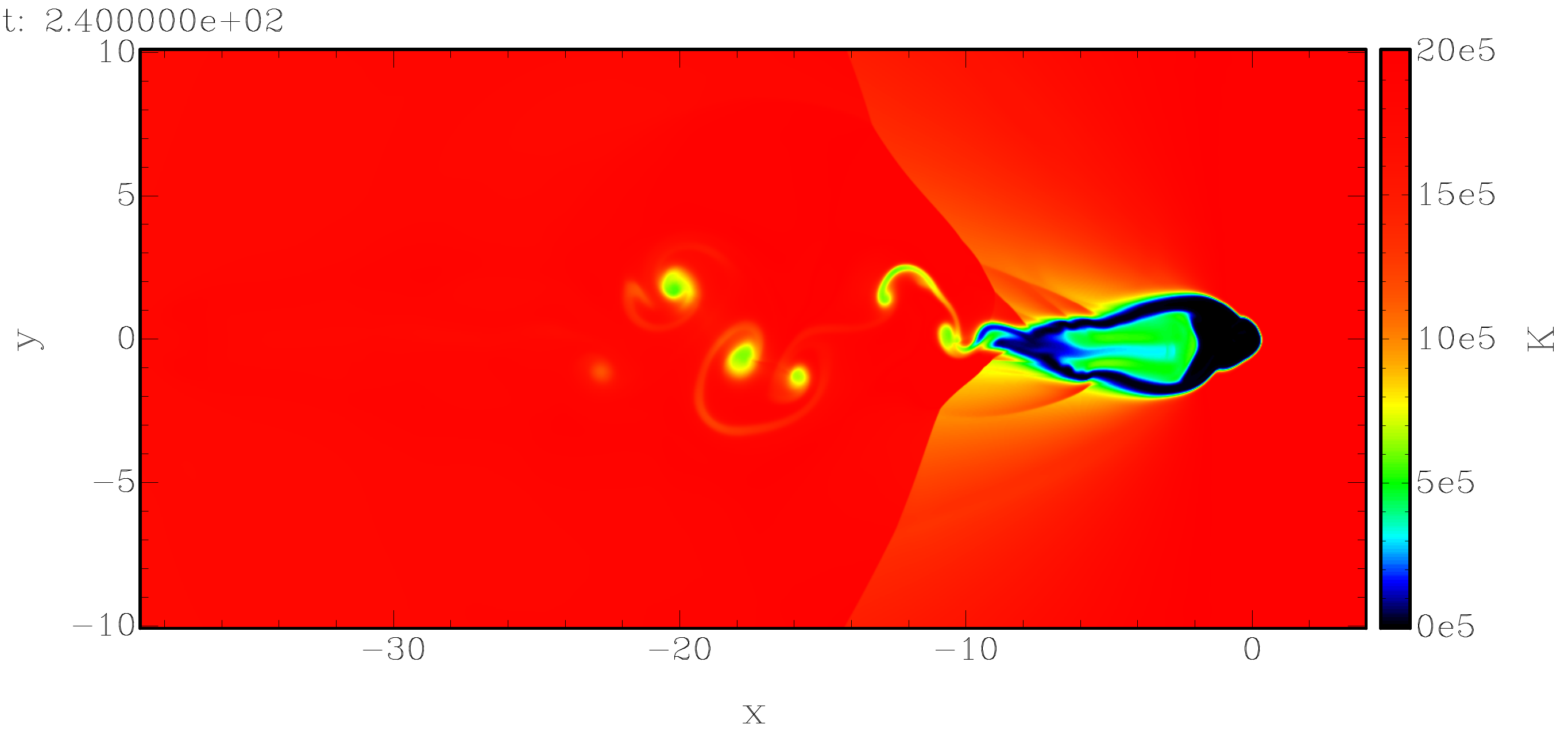}
\includegraphics[height=0.49\textwidth,angle=-90]{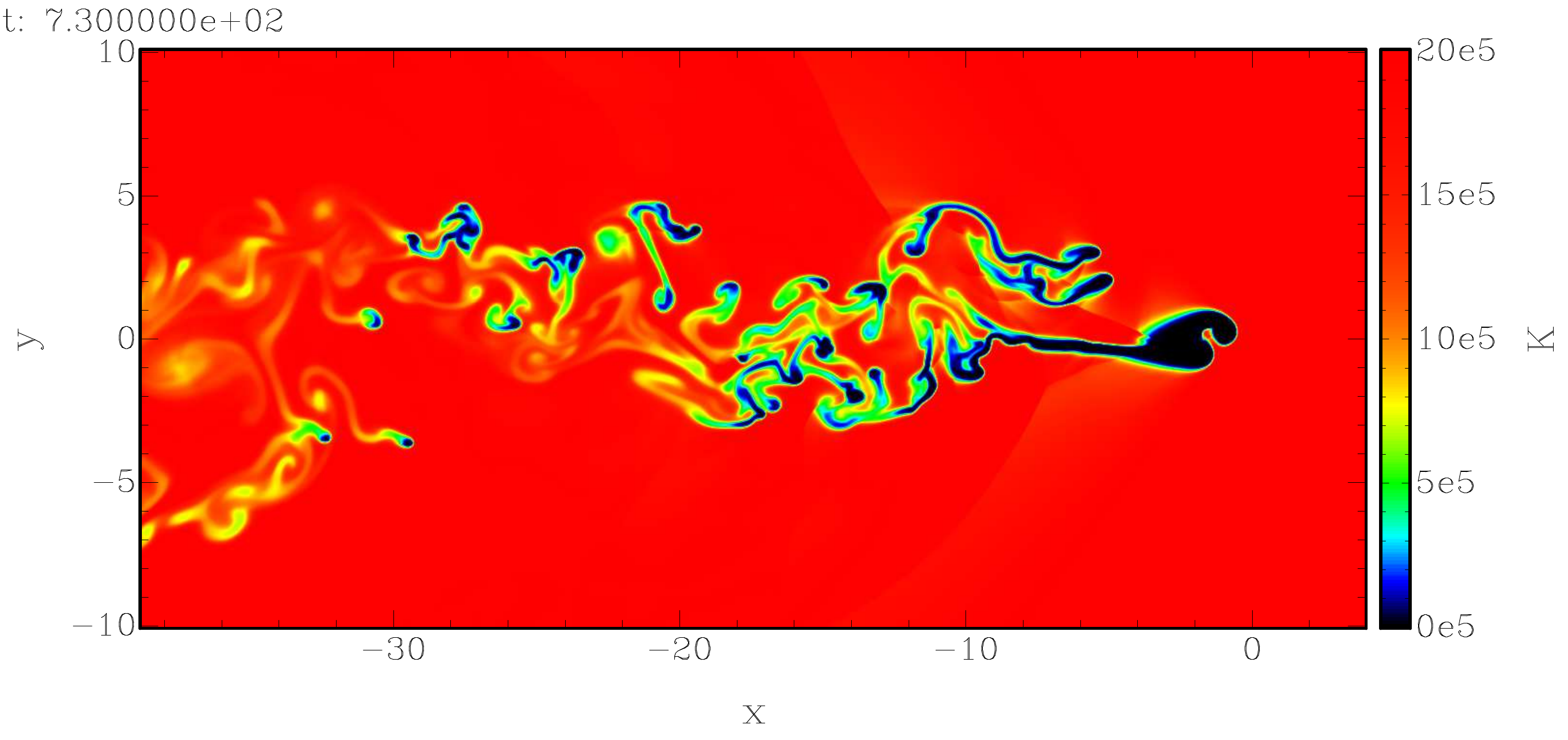}
\includegraphics[height=0.49\textwidth,angle=-90]{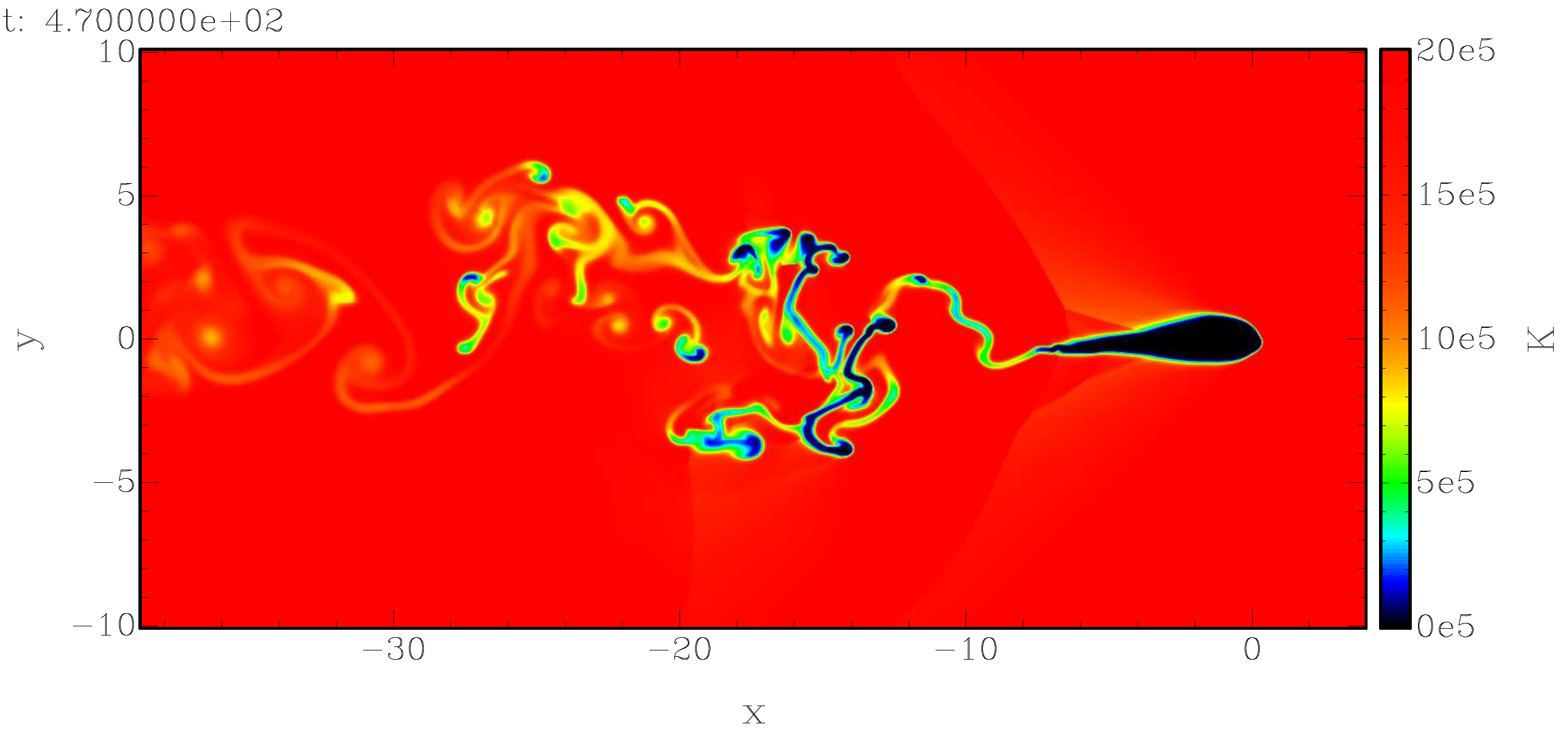}
\includegraphics[height=0.49\textwidth,angle=-90]{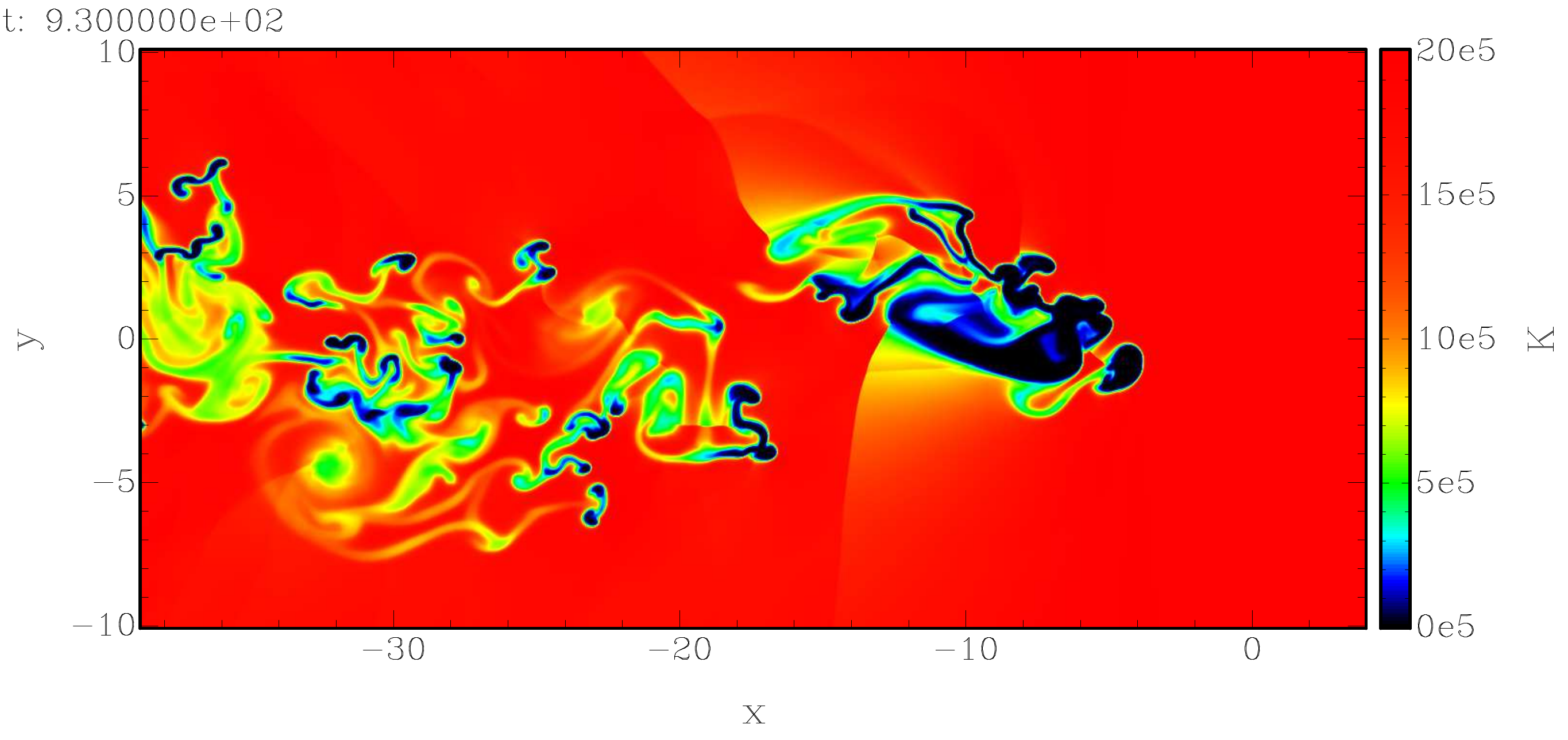}
\caption{Time evolution of the temperature distribution for our fiducial setup for the Sextans
dSph with $\ncor=1.8\times10^{-4}\cm$. In the left column (top to bottom) we show $t=$ 0, 240, and 470 Myr, and
in the right column we plot $t=$ 700, 730, and 930 Myr. The bottom left panel corresponds 
roughly to the time of pericentric passage. We only show a small section of the box. 
The axes are given in kpc.}
\label{fig:simSextans}
\end{figure*}
In \S \ref{sec:simulationsSex}, we describe our fiducial simulation setup for Sextans,
which has been obtained by taking the orbit with the median value of
the pericentric distance $\bar{r}_{\rm p}$. For this fiducial setup, we
illustrate the principal results of our analysis, in particular the procedure
that we adopted to determine the coronal density (averaged over the distance
range encompassed by the orbit) that produces complete stripping of the dwarf's
ISM. We examine, in \S \ref{sec:densitySex}, how the estimate for the
coronal density is affected by the choice of the orbit and the uncertainties in
the initial conditions. In \S \ref{sec:densityCar}, we compare the
values for the coronal density that we infer from Carina's simulations with
those found for Sextans, and in \S \ref{sec:temperatures} we show how
the choice of different temperatures for the coronal gas affects the results.

\subsection{Ram-pressure stripping from Sextans}\label{sec:simulationsSex}
We first examine the stripping of Sextans with the orbit parametrized by
$\bar{r}_{\rm p}=59.8 \kpc$ (see \S \ref{sec:orbits})
and all other parameters as quoted in Tables \ref{tab:properties} and
\ref{tab:initialConditions} and in the first line of Table \ref{tab:simsetup},
which represents our fiducial setup.
We then run a series of simulations varying only
the mean coronal density $\ncor$ until we find the value that produces
complete stripping of gas within the time of the simulation.
We find that the minimum coronal density needed for stripping 
to occur is $\ncormin=1.8\times10^{-4}\cm$.

\begin{figure}
\includegraphics[width=0.5\textwidth]{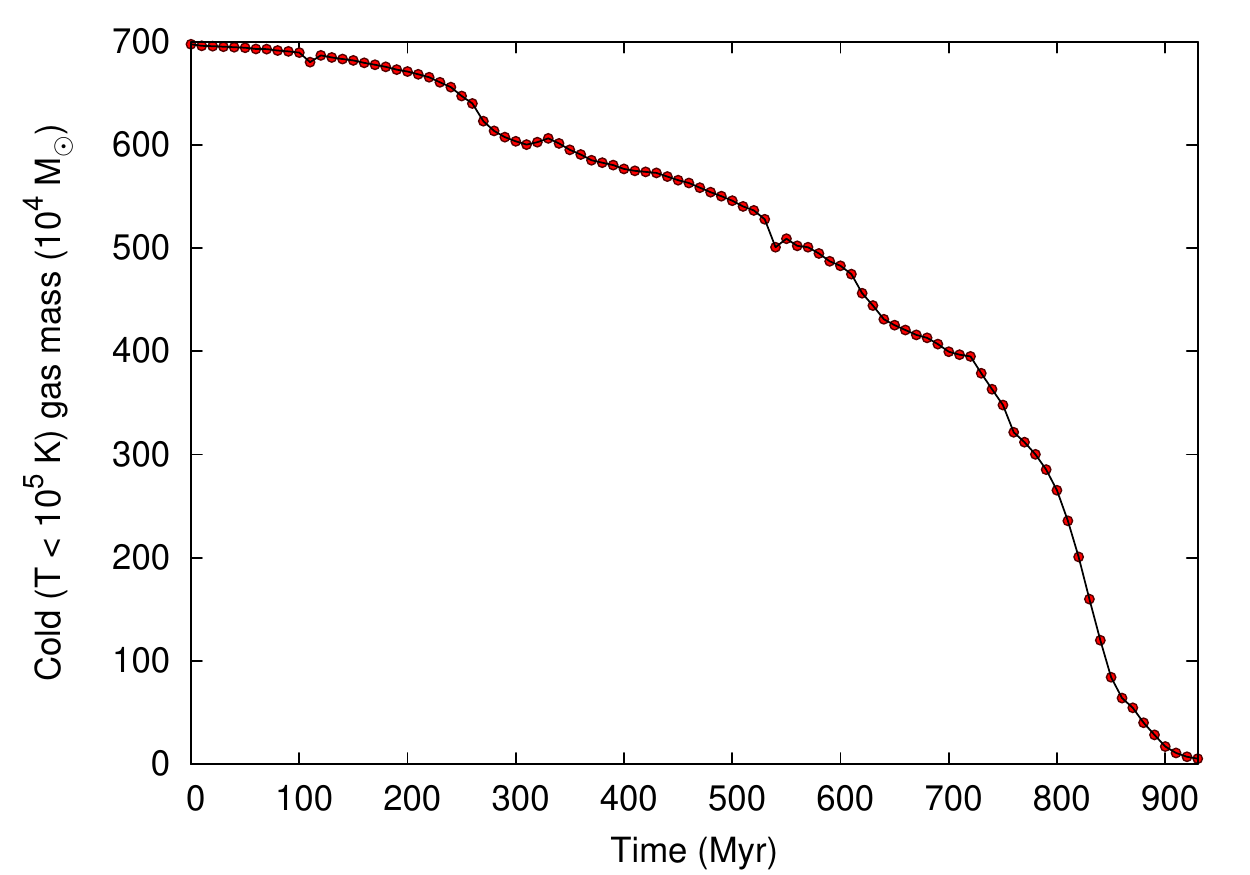}
\caption{Mass of cold ($<10^5$ K) gas gravitationally bound to the DM halo of
Sextans as a function of time from the beginning of our fiducial simulation.
The pericentre passage occurs at $t=465 \Myr$.}
\label{fig:boundMass}
\end{figure}

Fig. \ref{fig:simSextans} shows the temperature distribution at times
$t=0, 240, 470, 700, 730, 930$ Myr. 
We see that, as the dwarf galaxy starts to experience the
ram pressure exerted by the corona, a wake of stripped gas is formed. 
This wake becomes progressively more elongated and structured as time passes.
In this wake knots of cold gas ($T \sim
10^4$\,K) and regions at intermediate temperatures ($\sim 10^5$\,K) co-exist. The
presence of these intermediate temperature regions is indicative of a mixing
between the stripped dwarf's ISM and the coronal material. The gas removal
is not an instantaneous process. The mass loss rate is initially rather low and
increases after the dwarf has passed the orbit pericentre.
In Fig. \ref{fig:boundMass}, we
show the evolution of the mass of the cold gas mass bound to Sextans
(i.e. all gas with velocity less than the local escape velocity and $T<10^5\,\K$). 
The mass of bound gas decreases steadily and at an increasing rate
throughout the simulation. The increasing mass loss rate is a result of
the progressive disruption of the dwarf by ram-pressure stripping assisted by SN feedback.
Before the pericentre, only roughly 20\% of the gas is lost, the other 80\% is lost in the second half of
the simulation. Approximately $1$\,Gyr is required to reach a final mass of cold, bound gas of
$\sim 5 \times 10^4\rm\mo$, $\sim 1$\% of the initial one.

\subsection{Coronal gas density: lower bounds, errors and upper bounds}\label{sec:densitySex}
To reliably estimate the MW's coronal density, different sets of initial conditions must be explored
to account for various uncertainties. The main model uncertainties are due to the orbit reconstruction, the
determination of the SFH and the star formation law. In this section, we consider in turn each of them.

\begin{figure}
\includegraphics[width=0.5\textwidth]{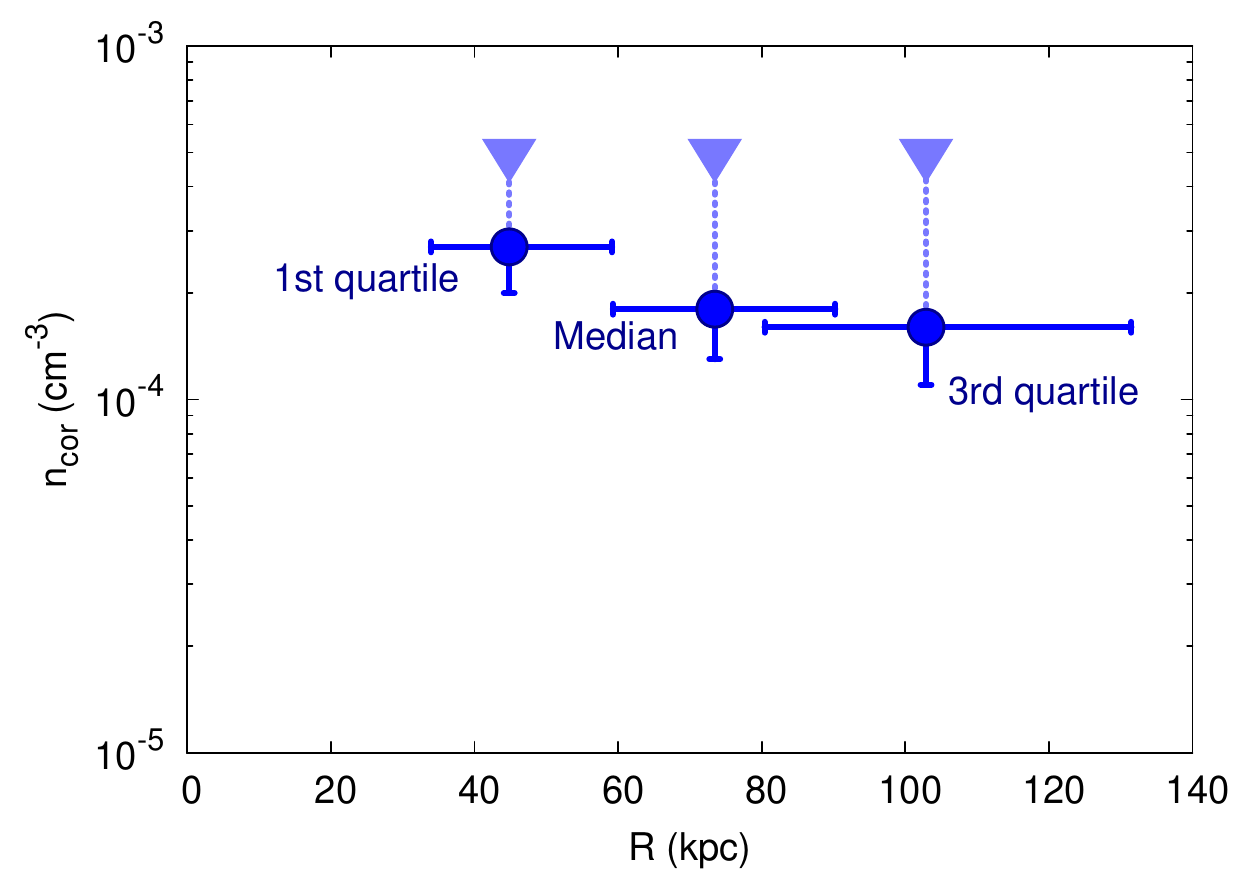}
\caption{Density of the corona of the MW that produces complete
gas stripping from the Sextans dSph.
The different determinations refer to three
representative orbits for the dSph with different pericentric radii, i.e. the
median orbit, and the first and third quartiles of the distribution of
pericentric radii. The down-pointing triangles show upper limits referred to
that specific radius. The derivation of errors and upper limits is described in
the text (\S \ref{sec:densitySex}).} 
\label{fig:coronaDensitySex}
\end{figure}

We start with the uncertainties in the orbit determination. Fig.
\ref{fig:coronaDensitySex} shows the minimum values of the density of the MW's
corona (points) that produce complete stripping from Sextans for the three representative
orbits chosen in \S \ref{sec:orbits}, i.e.\ the median value of $r_{\rm p}$
and the first and third quartiles of its distribution.
The error bar in the radius represents the range over which the coronal density has to be considered average
(see Table \ref{tab:simsetup}, eighth column, rows 1, 3, and 5, labelled as ``mid"), while the derivation of the
lower errors and the upper limits to the coronal density is described below.
The orbital parameters used to derive the coronal densities shown in Fig.\ 
\ref{fig:coronaDensitySex} are quite different (see Fig.\ \ref{fig:orbits} and \S \ref{sec:orbits}).
Nevertheless, the density required for the stripping is
similar for the three orbits and shows a nice decreasing trend with the distance
from the MW. This shows that the value of the coronal density is not too
sensitive to the specific choice of the orbital parameters.
The resulting values for $\ncormin$ are reported in Table \ref{tab:simulations}
labelled as ``mid''.

Next we explore both the effect of the uncertainties on the measured SFH and on
the applied star formation relation (equation (\ref{eq:SFlaw})), which influence the
value of initial gas density of the dwarf, $n_{\rm 0, gas}$. To investigate the effect of a lower dwarf ISM
density, we run an additional set of simulations (labelled as ``low'' in Tables
\ref{tab:simsetup} and \ref{tab:simulations}). We derive the lower limit of the
initial dwarf ISM density from an SFR of $2.4\times 10^{-5} \moyr$,
corresponding to reducing the fiducial value of $4.6\times 10^{-5} \moyr$ by $1
\sigma$ (see Table \ref{tab:initialConditions}). We then use
equation (\ref{eq:SFlaw}) with the upper $+0.6$ error to recover the lower $n_{\rm 0,
gas}=0.18 \cm$, which is shown in rows 2, 4, 6 of Table \ref{tab:simsetup}. This gives a lower
boundary for the coronal density which lies about $1 \sigma$ below the fiducial value
$\ncor=1.8\times10^{-4}\cm$. These values represent the lower error bars in Fig.\
\ref{fig:coronaDensitySex} for the different orbits. 

The above gives us a {\it robust lower bound} on the hot corona density. As outlined in
\S\ref{sec:pconfine}, we additionally use pressure equilibrium to estimate an upper bound
$\ncormax$ by setting the gas truncation radius $r_{\rm gas}$ equal to the star formation
radius $r_\mathrm{SF}$. In doing this, we are neglecting any  conspicuous redistribution of
stars after $t_{\rm lb}$. We plot the resulting upper limits as downward-pointing triangles
in Fig.~\ref{fig:coronaDensitySex}. Given the (large) uncertainties -- particularly on the orbit
of Sextans -- it is quite remarkable that all the values of the coronal density derived here
appear to be consistent with one another.

\begin{table}
\centering
\begin{tabular}{lccccc}
\hline
    dSph  & $r_{\rm p}$ & $\Delta\,r$ & $v_{\rm p}$ & $\Delta\,t_{\rm lb}$ & $\ncormin$\\
    point  & (kpc) & (kpc) & (km/s) & (Gyr) & ($\cm$)\\
\hline
Sextans\\
\hline
mid & 59.8 & 30.4 & 270.4 & 0.93 & $1.8\times10^{-4}$ \\ 
        low & 59.8 & 30.4 & 270.4 & 0.93 & $1.3\times10^{-4}$ \\
        mid & 33.9 & 25.3 & 333.6 & 0.42 & $2.7\times10^{-4}$ \\ 
        low & 33.9 & 25.3 & 333.6 & 0.42 & $2\times10^{-4}$ \\ 
        mid & 80.4 & 51.1 & 284.1 & 1.22 & $1.6\times10^{-4}$ \\ 
        low & 80.4 & 51.1 & 284.1 & 1.22 & $1.1\times10^{-4}$ \\
\hline
Carina\\
\hline  
mid & 51.2 & 30.6 & 291.4 & 0.74 & $1.7\times10^{-4}$ \\ 
        low & 51.2 & 30.6 & 291.4 & 0.74 & $1.5\times10^{-4}$ \\
\hline
\end{tabular}
\caption{Simulations that produced the complete stripping of gas from the dSphs.
The labels ``mid'' and ``low'' refer to the initial density of the dwarf's ISM.
$r_{\rm p}$ is the pericentre distance of the simulations, $\Delta\,r$ is the
considered spatial range from the pericentre found considering a stripping
efficiency greater than 50\% (see also Table \ref{tab:simsetup}), $v_{\rm p}$ is
the velocity at the pericentre, $\Delta\,t_{\rm lb}$ is the simulation time and
$\ncormin$ is the inferred {\it minimum} average coronal density
needed for stripping.}
\label{tab:simulations}
\end{table}

\subsection{Carina}\label{sec:densityCar}
We carry out a comparable set of simulations for the Carina dSph. For this dwarf
we use only one orbit, i.e. that with median $r_{\rm p}=51.2$ kpc, for which we
find $\ncormin=1.7\times10^{-4} \cm$. 
Estimating the lower error as before brings the lower limit down to
$\ncormin=1.5\times10^{-4} \cm$ (Table \ref{tab:simulations}).

\begin{figure}
\includegraphics[width=0.5\textwidth]{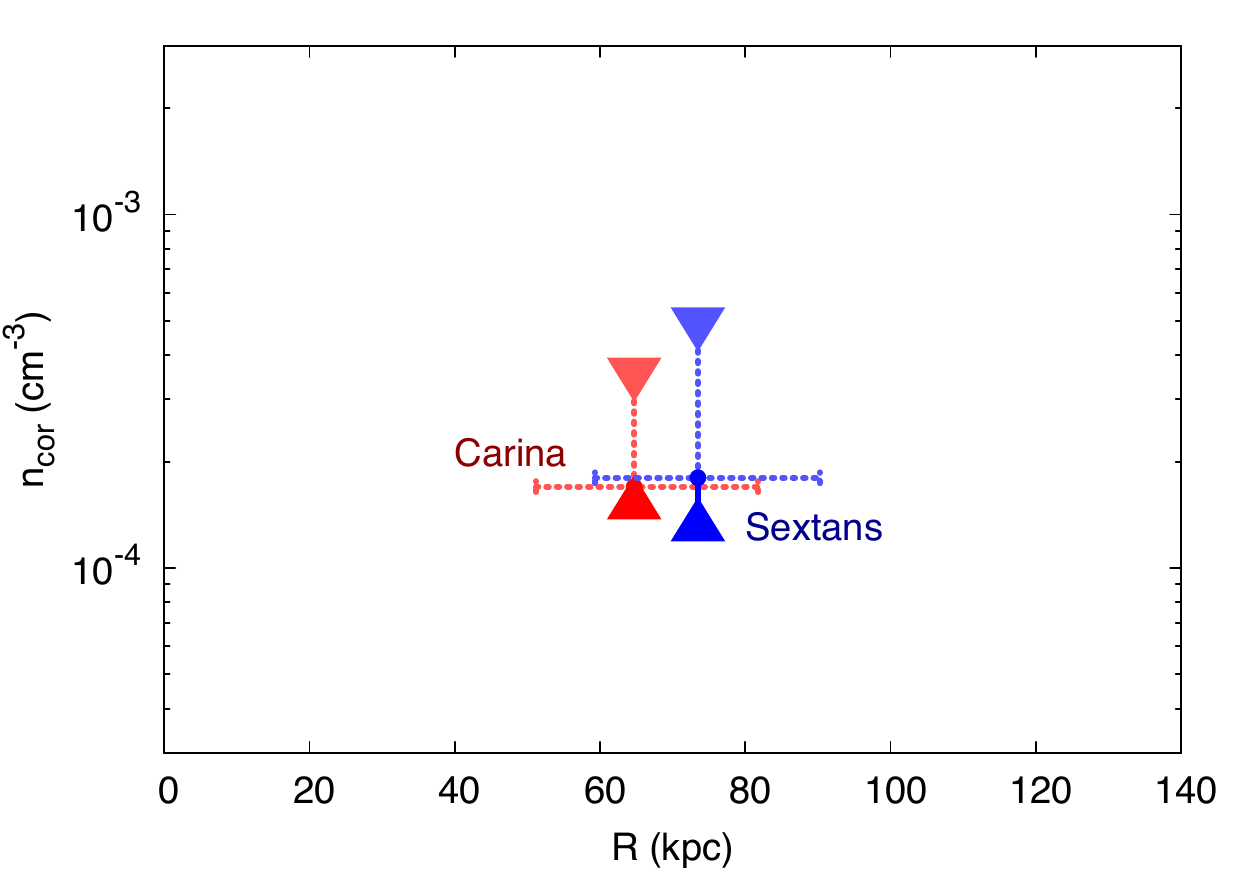}
\caption{Ranges of gas densities of the MW's corona allowed by
the Sextans (blue) and Carina (red) dwarfs. 
The derivation of the lower and upper bounds (triangles) is 
described in the text (\S \ref{sec:densitySex}).}
\label{fig:coronaDensity}
\end{figure}

We compare the coronal densities of the MW derived by using the median
orbit of Sextans and Carina in Fig. \ref{fig:coronaDensity}. As in Fig.
\ref{fig:coronaDensitySex}, the horizontal bar represents the range in radii
that we have considered for the simulation.
We now use upwards pointing triangles placed at the location of the lower $1\sigma$ error to denote our lower bound.
The upper limits (downwards pointing triangles) are estimated with the method described in \S \ref{sec:densitySex}.
The derived values for the coronal density are reported in Table
\ref{tab:simulations}, while in Table \ref{tab:results} we list the 
ranges of the radii and the upper and lower bounds of the coronal density
obtained for the median orbits of Sextans and Carina.
The two dSphs have rather different structural properties and orbital
parameters (see again Tables \ref{tab:properties}, \ref{tab:initialConditions}
and \ref{tab:simsetup}) and yet there is a remarkable consistency for the
recovered density values in the range of radii in which the two orbits overlap.
This fact further supports the basic soundness of the methodology that we adopt
here. Note in particular that the times of the last stripping ($t_{\rm lb}$) for
the two dwarfs are very different.
This may be an indication that the density of the Galactic corona has not 
changed significantly in the last $\sim 7 \Gyr$.
Note also that the assumption that there has been no significant redistribution of
the stellar component within the dwarf should be fully justified for Carina where 
$t_{\rm lb}$ is only $0.5 \Gyr$.

As a final end-to-end test of our systematic error, we consider a pericentric
passage at the peak of the SFH of Carina. By matching $\rho_\mathrm{gas}$ with
the value extracted from the following bin of the SFH, we derive a Galactic
corona density approximately three times larger than using $t_{\rm lb}$. It is 
possible that this systematic shift implies some evolution in Carina's orbit
over time; this interpretation will be considered in more detail in a separate
forthcoming paper. Here, we simply note that even this extreme test results in a
systematic error that is comparable to our other uncertainties.
\begin{table}
\centering
\begin{tabular}{cccc}
\hline
    Radius & range & $\ncormin$ & $\ncormax$\\
    (kpc) &  (kpc) & ($\cm$) & ($\cm$)  \\
\hline
 73.5   & 59.8-90.2 & $1.3\times 10^{-4}$ & $5\times 10^{-4}$  \\
 64.7   & 51.2-81.8 & $1.5\times 10^{-4}$ & $3.6\times 10^{-4}$  \\
\hline
\end{tabular}
\caption{Average density of the MW corona together with its upper and lower
limits as derived from the ram-pressure stripping along the median orbits
of Sextans and Carina.} 
\label{tab:results}
\end{table}

\subsection{Varying the coronal temperature}\label{sec:temperatures}
One of the main assumptions of our investigation is the temperature of the corona
at the location of the dwarf galaxies. To study the effect of different coronal
temperatures we run additional simulations using the median orbit of Carina with
the same parameters used before but different $T_{\rm cor}$. In particular, we
explore two additional coronal temperatures at $T_{\rm cor}=3\times 10^{6}$
and $1\times 10^{6}$\,K. The corresponding results, averaged over the range $51
<r<82 \kpc$ from the MW, are $\ncorminthree=1.5\times 10^{-4} \cm$ and $\ncorminone=2.5\times 10^{-4} \cm$. The increase (decrease) of the coronal
temperature causes the density to be lower (higher) than our fiducial value. In
\S\ref{sec:missingBaryons}, we discuss the implications of these results for
the missing baryon problem.

\section{Discussion}\label{sec:discussion}

\subsection{Missing baryons and the MW's corona}\label{sec:missingBaryons}
The results for the coronal density necessary for the stripping of gas from Sextans and Carina are summarized in Table \ref{tab:results}.
Here we show a conservative lower bound (fiducial value $-$ lower error) and the upper bound determined in \S\ref{sec:pconfine}.
We conclude that the coronal density, being a monotonic decreasing function of R, averaged between 50 and $90\kpc$ must be in the range $1.3\times 10^{-4} <n_{\rm cor}< 3.6\times 10^{-4} \cm$,
consistent both with the detection claims by \citet{Gupta+12} and with the analytical estimates of
\citet{Grcevich&Putman09}. We recall that $n_{\rm cor}$ is the total gas
density: $n_{\rm i} + n_{\rm e}$.
The lower limit is computed by subtracting the average $1\sigma$ of Sextans and Carina lower values to the value of the
coronal density ($n_{\rm cor}=1.75\times 10^{-4} \cm$) determined by averaging our 
fiducial ``mid'' simulations (Table \ref{tab:simulations}).
%The upper error is such that the upper limit of Carina, i.e.,
%the more stringent of the two, occurs at $\sim 3\sigma$ from $n_{\rm cor}=1.75$.
%} 
\begin{figure*}
\includegraphics[width=1.\textwidth]{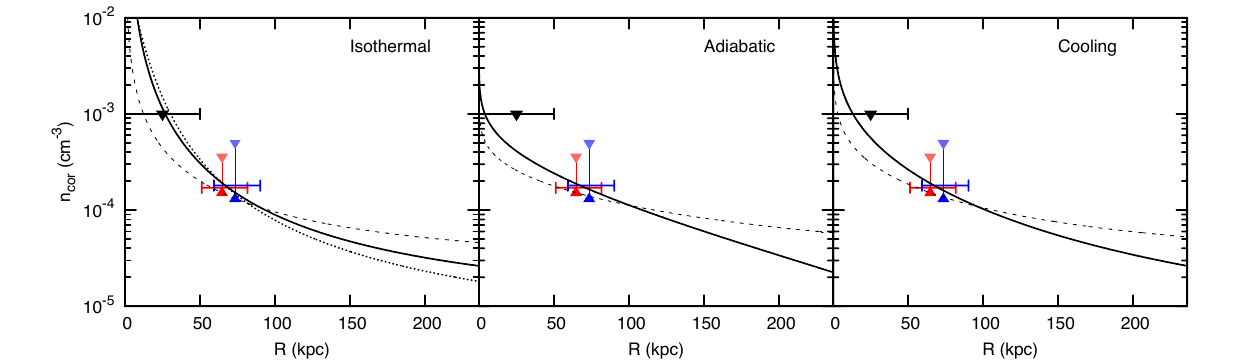}
\caption{Density profiles for different coronal models consistent with
all constraints: the range of coronal densities allowed by our analysis for Sextans and Carina, pulsar dispersion measures 
(black triangle) and X-ray emission upper limits (not shown). The solid line
corresponds to $T_\mathrm{cor}(50-90\,\kpc)=1.8\times10^6$\,K and the dashed line
to $T_\mathrm{cor}(50-80\,\kpc)=3\times10^6$\,K. The dotted line in the left-hand
panel shows results for an NFW potential (as opposed to our default TF
potential). From left to right, panels consider an isothermal ($\gamma = 1$),
adiabatic ($\gamma = 5/3$), and `cooling' ($\gamma = 1.33$) halo.} 
\label{fig:final}
\end{figure*}
\begin{table*}
\centering
\begin{tabular}{cccccc}
\hline
Potential & $T_{\rm cor}$ & \multicolumn{2}{c}{$\frac{M(<r_\mathrm{vir})}{M_\mathrm{mb}}$} & \multicolumn{2}{c}{$\frac{M(<r_\mathrm{vir})}{\mo}$}\\
- & ($10^6 \K$) & Isothermal & Adiabatic & Isothermal & Adiabatic\\
\hline
TF  &1.8 &$15-20$ \% &$16-48$ \% &$3.6\times 10^{10} - 4.8\times 10^{10}$ &$3.8\times 10^{10} - 1.1\times 10^{11}$\\
NFW &1.8 &$11$ \% &$9-25$ \%  &$3.4\times 10^{10}$ &$2.8\times 10^{10} - 7.7\times 10^{10}$\\
TF &3.0 & $22-50$ \% &$26-100$ \%  &$5.3\times 10^{10} - 1.2\times 10^{11}$ &$6.2\times 10^{10} - 2.4\times 10^{11}$\\
NFW &3.0 &$16-33$ \% &$18-74$ \%  &$4.5\times 10^{10} - 10^{11}$ &$5.6\times 10^{10} - 2.3\times 10^{11}$\\  
\hline
\end{tabular}
\caption{The fraction and mass of the missing baryons contained in coronae for 
different combinations of coronal temperature, equation of state and
Galactic potential consistent with the observational constraints.}
\label{tab:mbfrac}
\end{table*}

It is possible to use our derived range of $n_{\rm cor}$
as a constraint for the \textit{global} density profile of the MW's
corona. This profile is obtained by following the procedure outlined in
\S\ref{sec:adiiso}. From the density profile one can extrapolate the total mass
of the corona within the virial radius of the MW, which can then be
compared to the missing baryonic mass of the Galaxy. In addition to $n_{\rm
cor}$, we take also into account two further constraints discussed in AB10:

\begin{enumerate} 
\item the dispersion measures along the line of sight to Large Magellanic Cloud (LMC) pulsars, from which AB10
 estimate an upper bound for the coronal density of $\overline{n}_{\mathrm{e}}=5\times10^{-4} \cm$ 
 averaged over 50\,kpc from the Galactic Centre; 
\item the upper limit for X-ray emission measure, assuming our fiducial value of the 
      coronal metallicity of $0.1\ Z_{\odot}$.
\end{enumerate} 
The dispersion measure of LMC pulsars is the more stringent constraint at present due 
to our assumption of a low coronal metallicity. 

Through equation (\ref{eqn:rhodist}), we compute a series of coronal density profiles
consistent with all of the above constraints, using three different assumptions
about the thermodynamic state of the coronal gas: isothermal ($\gamma = 1$),
adiabatic ($\gamma = 5/3$), and 'cooling' ($\gamma = 1.33$). As for the DM
potential we make two different choices: our default TF potential (see
equation (\ref{eqn:rhodist}) truncated at\footnote{This inner truncation is used to
avoid any contamination from the disc and does not affect the value of the
recovered coronal mass (see also AB10).} $10\,\kpc \leqslant R \leqslant R_{\rm
vir}=236$\,kpc, with $M_{\rm vir}=1.54\times 10^{12} \mo$); and an NFW profile.
We also consider three different coronal temperatures: $1.8\times 10^6$, 
$10^6$ and $3\times 10^6$\,K. The exploration of the parameter space
resulted in 21 models compatible with all of the constraints considered here.
In particular, we find that, regardless of the choice of the potential, for
the isothermal models our upper limits are less stringent than the constraint from the dispersion measure,
while for the adiabatic and cooling coronae they are roughly coincident.
Hence, the upper limits on the MW's baryon fraction described in this section are determined
by the dispersion-measure limit rather than our pressure-confinement method described in 
\S \ref{sec:pconfine}.

To derive the expected mass of the missing baryons $M_{\rm mb}$ associated with the Milky
Way we follow again AB10 and set $M_\mathrm{mb} = 15\%\ M_\mathrm{tot}\ $, where
$M_\mathrm{tot}$ is the sum of the DM mass ($M_\mathrm{vir}$) and the observed
baryons mass. For the latter we take $M_\mathrm{ob}=6\times
10^{10} \mo$ (see \S \ref{sec:intro}). The expected missing baryon mass is then
$2.4\times 10^{11} \mo$ for the TF model.

Our results are presented in Fig. \ref{fig:final}, where we show the ranges
given by Sextans and Carina and the coronal profiles with reference values $n_{\rm cor}=1.75\times 10^{-4}$ and $1.5\times 10^{-4} \cm$ for 
$T_{\rm cor} = 1.8 \times 10^6$ and $3 \times 10^6$\,K, respectively.
All the models with $T_{\rm
cor} = 10^6$\,K yielded no solution consistent with all of the constraints and
therefore are not shown. In the isothermal case, we find for $T_{\rm cor} = 1.8 \times
10^6$\,K ($3 \times 10^6$\,K) a coronal baryon fraction of 15-20\% (22-50\%) of
the expected MW's missing baryons, marginalizing over all uncertainties.
For adiabatic and `cooling' models instead, the temperature profile is no longer
constant and so our assumed coronal temperature corresponds to an average over
the ranges $50-90$ and $50-80\,\kpc$ for $T_{\rm cor} = 1.8 \times 10^6$ and 
$3 \times 10^6$\,K, respectively. The results for an
adiabatic or `cooling' halo are nearly indistinguishable, with a difference of
$\lesssim$\,2\% in the recovered missing baryon fractions. For $T_{\rm
cor}(50-90\,\kpc) = 1.8 \times 10^6$\,K, we find a coronal baryon fraction of
16-48\% of the expected missing baryons, while for $T_{\rm cor}(50-80\,\kpc) = 3
\times 10^6$\,K the value is 26-100\%. 
As expected, for an adiabatic or
`cooling' corona the baryon fraction can be significantly larger than in the
isothermal case the density in such a corona drops less rapidly with radius, allowing
more gas to be stored in the huge volume just inside the virial radius. An
adiabatic corona at high temperature could, in principle, contain all of the
MW's expected missing baryonic mass. These results are broadly consistent
with those of AB10 and also the estimates of missing baryon fractions in
external galaxies \citep{Anderson&Bregman11,Dai+12,Anderson+13}, although our
fractions are higher than theirs probably due to our lower value of the coronal
metallicity.

Finally, we consider how our assumption of a TF profile affects these results.
We use instead the NFW potential from AB10 with $R_{\rm vir}=250$\,kpc, $M_{\rm
vir}=2\times 10^{12} \mo$ and concentration parameter $c=12$ (see the dotted line,
left-hand panel of Fig. \ref{fig:final}). Notice that for an NFW profile, the gas
density falls more steeply leading to a lower extrapolated total mass. However,
the effect is typically quite small compared to the other
uncertainties. Our results for the missing baryon fractions are summarized in
Table \ref{tab:mbfrac}.
All isothermal models predict an amount of missing baryons in the corona
between 10 and 50\% of the expected value \citep[see also][]{MillerBregman13}.
If the hot gas has an adiabatic equation of state, the corona can accommodate more
gas and we can not rule out that it could contain the whole predicted amount of missing 
baryons \citep[see also][]{Fang+12}.

\subsection{Comparison with the analytic ram-pressure stripping formula}\label{sec:compAnalytic}
\begin{table}
\centering
\begin{tabular}{lcccc}
\hline
    Point & $n_{\rm an}$ & $n_{\rm sim}$ & $\overline{n}_\mathrm{gas}$ & $\overline{v}_\mathrm{x}$\\
          &  ($\cm$)     & ($\cm$)       & ($\cm$)            & ($\kms$) \\
\hline
Sextans\\
\hline
Median & $3.6\times 10^{-5}$ & $1.8\times 10^{-4}$ & $0.09$ & $228$ \\
First quartile & $2.3\times 10^{-5}$ & $2.7\times 10^{-4}$ & $0.09$ & $286$ \\
Third quartile & $3.1\times 10^{-5}$ & $1.6\times 10^{-4}$ & $0.09$ & $246$ \\
\hline
Carina\\
\hline
Median & $3.2\times 10^{-5}$ & $1.7\times 10^{-4}$ & $0.14$ & $251$\\
\hline
\end{tabular}
\caption{Comparison between the predicted values of the minimum coronal density for stripping from equation (\ref{eq:eqan}) (column 1) and the results of the ``mid'' simulations (column 2) listed in Tables \ref{tab:simsetup} and \ref{tab:simulations} (see the text).
}
\label{tab:analytic}
\end{table}
We now compare the results of our simulations (Table \ref{tab:simulations}) with
the analytic estimates computed from equation (\ref{eqn:stripformperi}). As in our
simulations we approximate the motion of the dwarf through the corona as
one-dimensional, the minimum coronal density averaged in the distance
range $\Delta r$ near the pericentre required to completely strip the gas away is 
\begin{equation}
\label{eq:eqan}
\ncormin(\Delta r) \sim \dfrac{\sigma_\mathrm{x}^2\ \overline{n}_\mathrm{gas}}{\overline{v}_\mathrm{x}(\Delta r)^2}\ ,
\end{equation}
where $\overline{n}_\mathrm{gas}$ is the average gas density of the dwarf within
$r_\mathrm{gas}$, $\overline{v}_\mathrm{x}(\Delta r)$ the average one-dimensional velocity of the
dwarf ($v_\mathrm{x}(\Delta r)$=$v_\mathrm{r}(\Delta r)$ in our
simulations) and $\sigma_\mathrm{x}$ the $x$-component of the central,
isotropic stellar velocity dispersion.
Table \ref{tab:analytic} presents our findings for the ``mid'' (fiducial) simulations
listed in Tables \ref{tab:simsetup} and \ref{tab:simulations}. The analytic estimates
have been obtained by considering $\sigma=7.9$ and $6.6\ \kms$ for Sextans and Carina, respectively
\citep{Walker+09}. Our numerical results are greater by about a factor of 5 with respect to the analytic predictions of equation (\ref{eq:eqan}). For the first quartile orbit of Sextans, this difference reaches a factor of 10.
We can conclude that over the small range of densities used here, the analytic formula for
stripping does not give a fairly good estimate of the coronal density, leading to the conclusion that
non-linear effects are fundamental in recovering realistic values of $\ncormin$.

\subsection{SN feedback}\label{sec:SNe}
\begin{figure}
\includegraphics[width=0.5\textwidth]{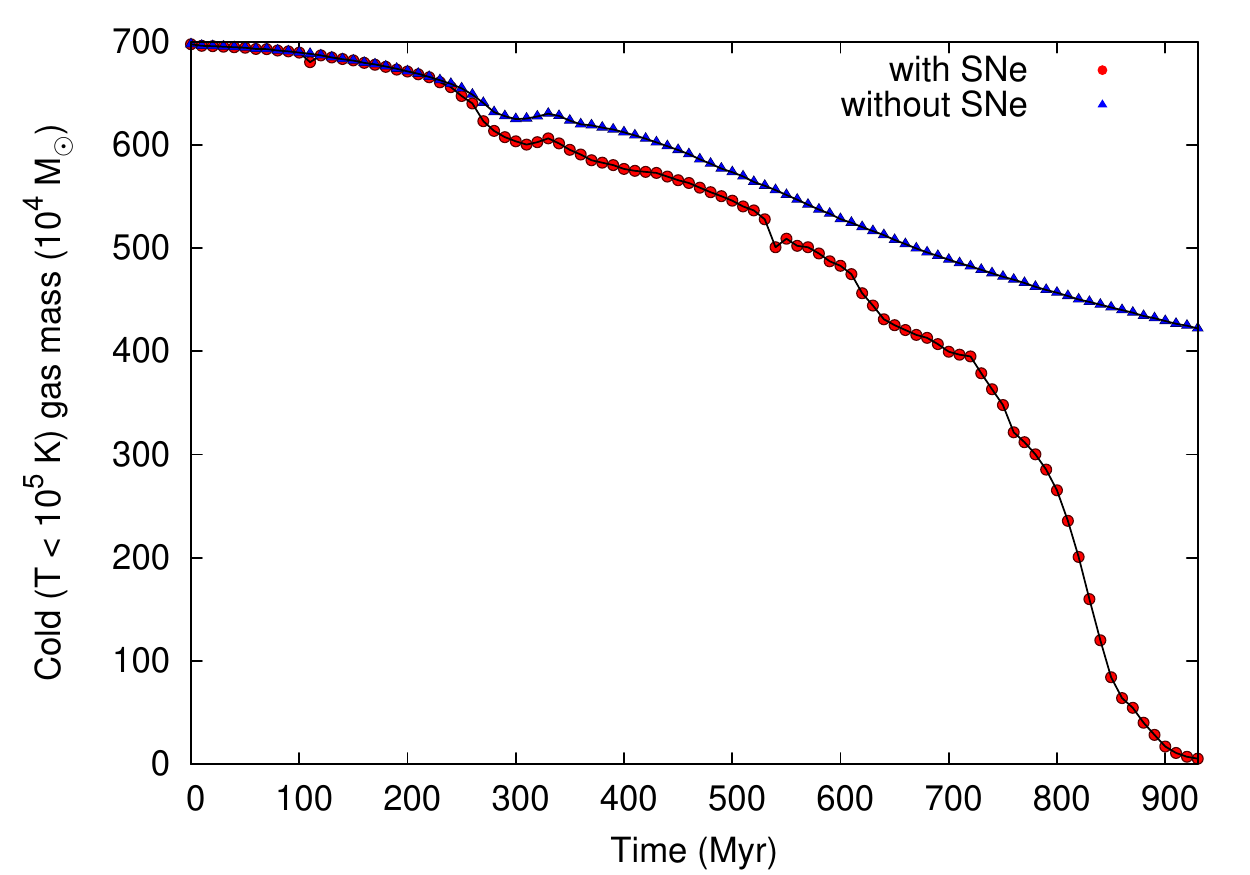}
\caption{Cold, bound gas mass for the fiducial simulation of Sextans with and without SNe.}
\label{fig:w-oSNe}
\includegraphics[width=0.5\textwidth]{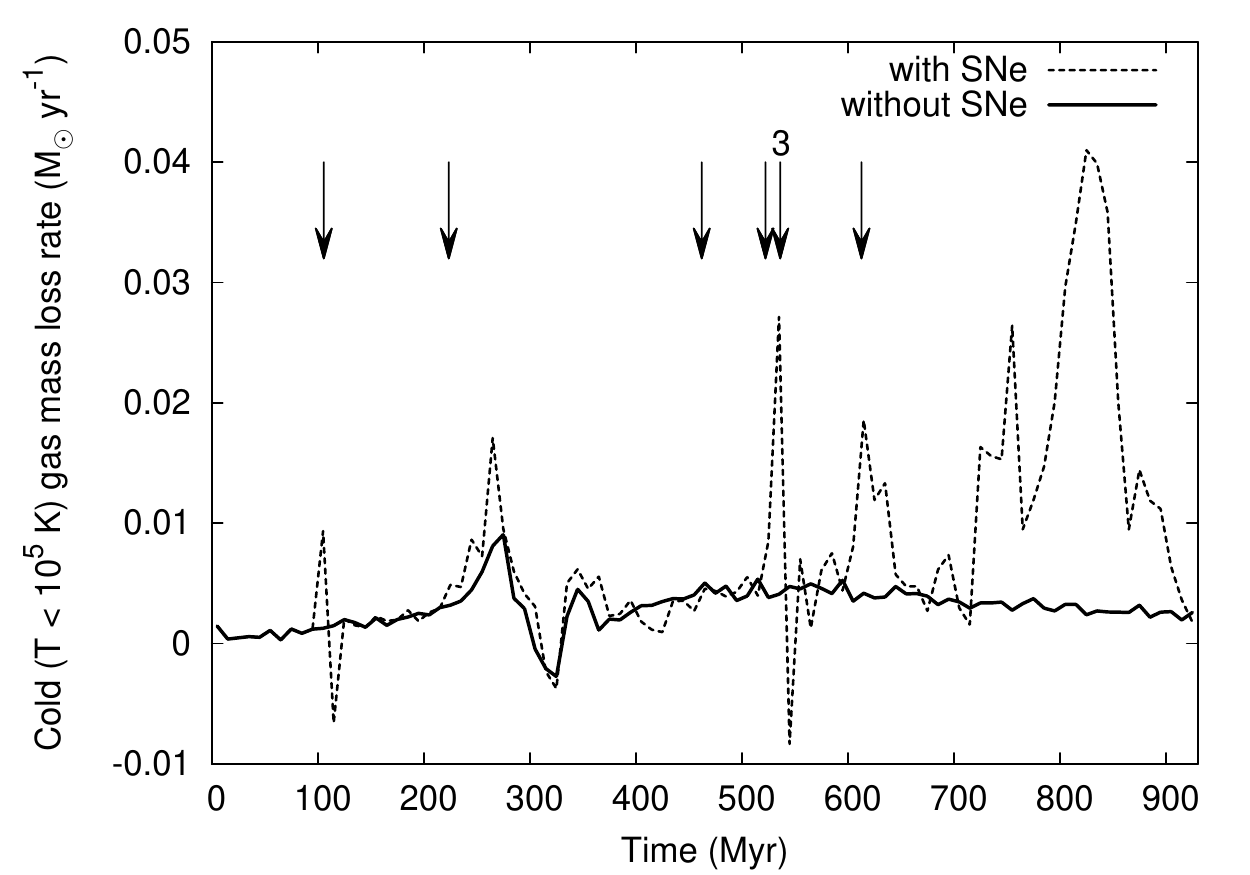}
\caption{Cold gas mass loss rates for the fiducial simulation of Sextans
with and without SNe. The arrows represent the time at which an SN has exploded;
the number 3 over the fifth arrow means that a burst has occurred (three SNe in
8 Myr).} 
\label{fig:rates}
\end{figure}
One of the novel features of this work is the introduction of discrete SN
injections. If we do not consider SN explosions, the stripping process should
naturally evolve towards a Kelvin-Helmholtz (KH) assisted regime. Additionally,
the fact that we are using a varying dwarf velocity means that we expect the
development of Rayleigh-Taylor (RT) instabilities. However, SN explosions are
very efficient at changing the local morphology of the gas distribution, leading
to an effective disruption of the RT/KH seeds. In practice, SNe destroy the
regular flow at the interface between the hot and cold gas, leading to a
SN-assisted stripping process. Without considering SN explosions, the gas flow
past the dwarf is rather smooth and eddies form. Including SNe, as shown in Fig.
\ref{fig:simSextans}, causes the flow to be quite clumpy. Perhaps, these cold
clumps travelling at a few hundreds of $\kms$ are related to (some of) the MW's
HVCs \citep[see also][]{Mayer+06, Binney+09}.

Figs \ref{fig:w-oSNe} and \ref{fig:rates} show the evolution of the cold
bound mass and the cold mass loss rate for the fiducial simulation of Sextans
(median orbit and $n_{\rm cor}=1.8\times 10^{-4} \cm$) with and without SNe. SN
explosions increase the cross-section of the cold gas distribution, leading to a
more efficient stripping and a mass loss rate that can become four times larger
than that without SNe (see Fig.\ \ref{fig:rates}). On the other hand, the same
simulation including SNe but without ram-pressure stripping leads to an
inefficient gas removal process, with a final cold gas mass very close to the
initial one. For this reason, we conclude that it is the {\it combination} of
SNe and ram pressure that is key for recovering the correct stripping rate.
Without SNe, the coronal density required to completely strip away the gas is
$\ncormin=2.9\times 10^{-4} \cm$ -- about two times higher than for our
reference simulation with SNe. This is higher than independent observational
limits on the coronal density (see \S \ref{sec:missingBaryons}), suggesting that
SN explosions are critical for recovering realistic coronal profiles \citep[see
also][]{Nichols&Bland-Hawthorn11}.

\section{Conclusions}\label{sec:conclusions}
We have performed a suite of hundreds of hydrodynamical simulations of
ram-pressure stripping of two dwarfs, Sextans and Carina, in order to estimate
a lower bound on the coronal gas density of the MW. 
In addition, we have derived an upper bound by considering the pressure confinement of these dwarfs by the hot corona. 
We have introduced several novel
features as compared to previous analyses: realistic orbits for the dwarfs, a
model of discrete SN feedback and a recovery of the initial gas mass contained
in the dwarfs (determined from their measured SFHs). We find
that the coronal number density in the range $50-90 \kpc$ from the Galaxy must be in
the range $ 1.3 \times 10^{-4} < n_\mathrm{cor} <  3.6 \times 10^{-4} \cm$. 
We have considered many sources of systematic and random error ensuring that this result is robust.
% within our quoted uncertainties. 

We have derived coronal models consistent with our lower and upper bounds on the 
coronal density, X-ray emission limits, and pulsar dispersion measures. The pulsar constraint is particularly important in providing a more rigorous upper bound on the coronal density than our pressure confinement calculation (that requires an additional assumption about the radial extent of star formation within the dSphs). We have explored different coronal temperatures, Galactic potentials and equations
of state for the gas, computing a set of coronal density profiles consistent
with all of the above constraints. Extrapolating the baryonic mass in these
models to large radii, we have estimated the fraction of `missing baryons' that
can exist in a hot corona within the MW's virial radius. Considering
as a reference model an isothermal corona at
$T_\mathrm{cor}=1.8\times10^6\K$ in hydrostatic equilibrium with the Galactic
potential, the missing baryon fraction is 10-20\%. Hotter and/or adiabatic coronae can contain more baryons than our reference
model. However, of the set of 21 coronal density profiles analysed 
in this work, only one model (hot and adiabatic) is consistent with all of
the expected missing baryons lying within the virial radius of the MW. 
Thus, models for the MW must either explain why its corona is in a hot, adiabatic thermal state, or why a large fraction of the MW's baryons either never fell in, or were removed by energetic feedback.

\section*{Acknowledgements}

We thank an anonymous referee for comments and suggestions that improved the clarity of this work.
We also thank E.\ Held and L.\ Rizzi for providing data and M.\ Anderson, G.\ Battaglia, J.\ Binney, T.\ Naab, and S.\ White for helpful comments and discussions.
We acknowledge the CINECA awards N. HP10CPPUNB and HP10C8MF3E (2011), the ETHZ
Brutus Cluster and the RZG Odin Cluster for the availability of high performance
computing resources and support. AG and SW thank the Deutsche
Forschungsgemeinschaft (DFG) for funding through the SPP 1573 ``The Physics of
the Interstellar Medium''. FF acknowledges support from PRIN-MIUR, project ``The Chemical and Dynamical Evolution of the Milky Way and Local Group Galaxies'', prot.\ 2011SPTACC.
JIR would like to acknowledge support from SNF grant
PP00P2\_128540/1. FM is supported by the collaborative research centre ``The
Milky Way System'' (SFB 881) of the DFG through subproject A1. 
HL acknowledges a fellowship from the
European Commissions Framework Programme 7, through the Marie Curie Initial
Training Network CosmoComp (PITN-GA-2009-238356). 
The original idea for this work arose during the inspiring
meeting ``Problematik'' that took place at Magdalen College, Oxford (UK), in
2010 August.

\bibliographystyle{mn2e}
\bibliography{UnvCorMWvRPsDwSats}

\label{lastpage}

\end{document}